\RequirePackage[orthodox, l2tabu]{nag}
\documentclass[12pt]{iopart}
\usepackage{iopams}
\expandafter\let\csname equation*\endcsname\relax
\expandafter\let\csname endequation*\endcsname\relax
\usepackage{amsmath}
\usepackage{siunitx}
\usepackage{graphicx}
\usepackage{epstopdf}
\usepackage{microtype}
\usepackage[dvipsnames]{xcolor}
\usepackage{hyperref}
\usepackage[figuresright]{rotating}
\begin{document}
\title[]{Bayesian inference of axisymmetric plasma equilibrium}

\author{Sehyun~Kwak$^{1}$, J.~Svensson$^{1}$, O.~Ford$^{1}$, L.~Appel$^{2}$, Y.-c.~Ghim$^{3}$ and JET~Contributors\footnote{See the author list of ‘Overview of JET results for optimising ITER operation’ by J. Mailloux et al. to be published in Nuclear Fusion Special issue: Overview and Summary Papers from the 28th Fusion Energy Conference (Nice, France, 10-15 May 2021)}}

\address{EUROfusion Consortium, JET, Culham Science Centre, Abingdon, OX14 3DB, United~Kingdom}
\address{$^{1}$Max-Planck-Institut f\"{u}r Plasmaphysik, 17491 Greifswald, Germany}
\address{$^{2}$Culham Centre for Fusion Energy, Culham Science Centre, Abingdon, OX14 3DB, United~Kingdom}
\address{$^{3}$Department of Nuclear and Quantum Engineering, KAIST, Daejeon 34141, South~Korea}

\ead{sehyun.kwak@ipp.mpg.de}

\vspace{10pt}
\begin{indented}
\item[]\today
\end{indented}

\begin{abstract}
We present a Bayesian method for inferring axisymmetric plasma equilibria from the magnetic field and plasma pressure measurements. The method calculates all possible solutions for plasma current and pressure distributions consistent with the measurements and magnetohydrodynamic (MHD) force balance. Toroidal plasma current and magnetic field coils are modelled as a set of axisymmetric current-carrying solid beams. The other parameters such as plasma pressure and poloidal current flux are given as a function of poloidal magnetic flux, which is determined given a 2D current distribution. Plasma pressure and poloidal current flux profiles are modelled as Gaussian processes whose smoothness is optimally chosen based on the principle of Occam's razor. To find equilibrium solutions, we introduce an MHD force balance constraint at every plasma current beam as a part of the prior knowledge. Given all these physical quantities, predictions calculated by the predictive (forward) models for diagnostics are compared to the observations. The high dimensional complex posterior probability distribution is explored by a new algorithm based on the Gibbs sampling scheme.
\end{abstract}

\vspace{2pc}

%\submitto{\NF}

%\maketitle

\section{Introduction}\label{sec:introduction}
One of the approaches to generating fusion power is to confine fusion fuel in the form of a plasma by using a magnetic field generated by external coils. In the magnetic field, the fusion plasma, which is an electrically conducting fluid-carrying internal currents, experiences a magnetic pressure due to the Lorentz force. This magnetic pressure balances out the plasma pressure and maintains the plasma in a magnetohydrodynamic (MHD) equilibrium state. Predicting this MHD equilibrium is critical for plasma control and physics studies \cite{Ferron1998, Wesson2004, Freidberg2008, Freidberg2010}. The equilibrium current distribution determines the magnetic field geometry of the fusion plasma that provides the canonical coordinate system, in which to express physical quantities for further research, for example, energy transport. In an axisymmetric fusion device like a tokamak, this magnetic field geometry can be represented as a set of poloidal magnetic flux surfaces often normalised to zero at the plasma centre, known as a magnetic axis, and to one at the plasma boundary, known as the last closed flux surface (LCFS) \cite{Wesson2004}.

The conventional way of inferring an equilibrium current distribution is to find a single solution to an MHD force balance equation such as the Grad-Shafranov equation \cite{Grad1958, Shafranov1963}. This equilibrium solution can be found iteratively by, for example, the equilibrium fitting (EFIT) code \cite{Lao1985}. This approach has been providing an equilibrium solution successfully, nevertheless, it has the following limitations: it often makes use of a simple 1D parameterisation of plasma current and pressure with a handful of parameters, which might underfit the data, and it may only take into account magnetic field measurements thus this equilibrium solution might be inconsistent with other data, for example, plasma electron density and temperature measurements. Moreover, this approach typically finds only a single solution, not all possible solutions which might explain the data within their predictive uncertainties. In other words, this conventional approach might not provide posterior uncertainties of the plasma equilibrium current (and pressure) distribution.

In this work, we present a Bayesian method for inferring axisymmetric plasma equilibria consistent with various data from multiple plasma diagnostics for the magnetic field, electron density and temperature measurements. This method is developed based on the current tomography method \cite{Svensson2008} and the equilibrium model \cite{Ford2010} in which toroidal plasma current and external magnetic field coils are modelled as a set of axisymmetric current-carrying solid beams. Given a 2D current distribution, we can determine poloidal magnetic flux surfaces. The other parameters such as plasma pressure and poloidal current flux are given as a function of poloidal magnetic flux and mapped to the 3D Cartesian coordinates. Given all these physical quantities, predictions calculated by the predictive (forward) models for the diagnostics are compared to the observations. The method is implemented for the Joint European Torus (JET) tokamak experiment and takes into account the magnetic probes (pickup coils, saddle coils and flux loops), polarimeters, interferometers, Thomson scattering and lithium beam emission spectroscopy systems. Although we have a substantial amount of data, it is not enough to infer all these physical quantities in this tomographic problem. Therefore we have to introduce our prior knowledge to exclude unreasonable solutions. We make use of non-parametric Gaussian processes to model plasma pressure and poloidal current flux profiles, and the smoothness of the profiles is optimally chosen based on the principle of Occam's razor \cite{Gull1988, MacKay1991, Svensson2011_GP}. In addition, to find equilibrium solutions, we implement an MHD force balance constraint at every plasma current beam by introducing so-called \textit{virtual observations} as a part of the prior knowledge. These virtual observations exclude non-equilibrium solutions in the parameter space, thus we get solutions that fulfil the MHD force balance \cite{Ford2010}. For comparison, we also perform inference without the equilibrium prior. The solutions are provided as the full joint posterior probability distribution of plasma current and pressure. However, exploring this high dimensional complex posterior distribution is computationally challenging \cite{Ford2010, VonNessi2013, VonNessi2014}. To overcome this problem, we developed a sampling algorithm based on the Gibbs sampling scheme \cite{Geman1984}. In short, the algorithm splits the full joint posterior distribution into a couple of low dimensional conditional posterior distributions and samples them consecutively. In this way, we can substantially reduce the difficulties of sampling the full joint posterior distribution.

\begin{figure}
	\centering
	\includegraphics[width=\linewidth]{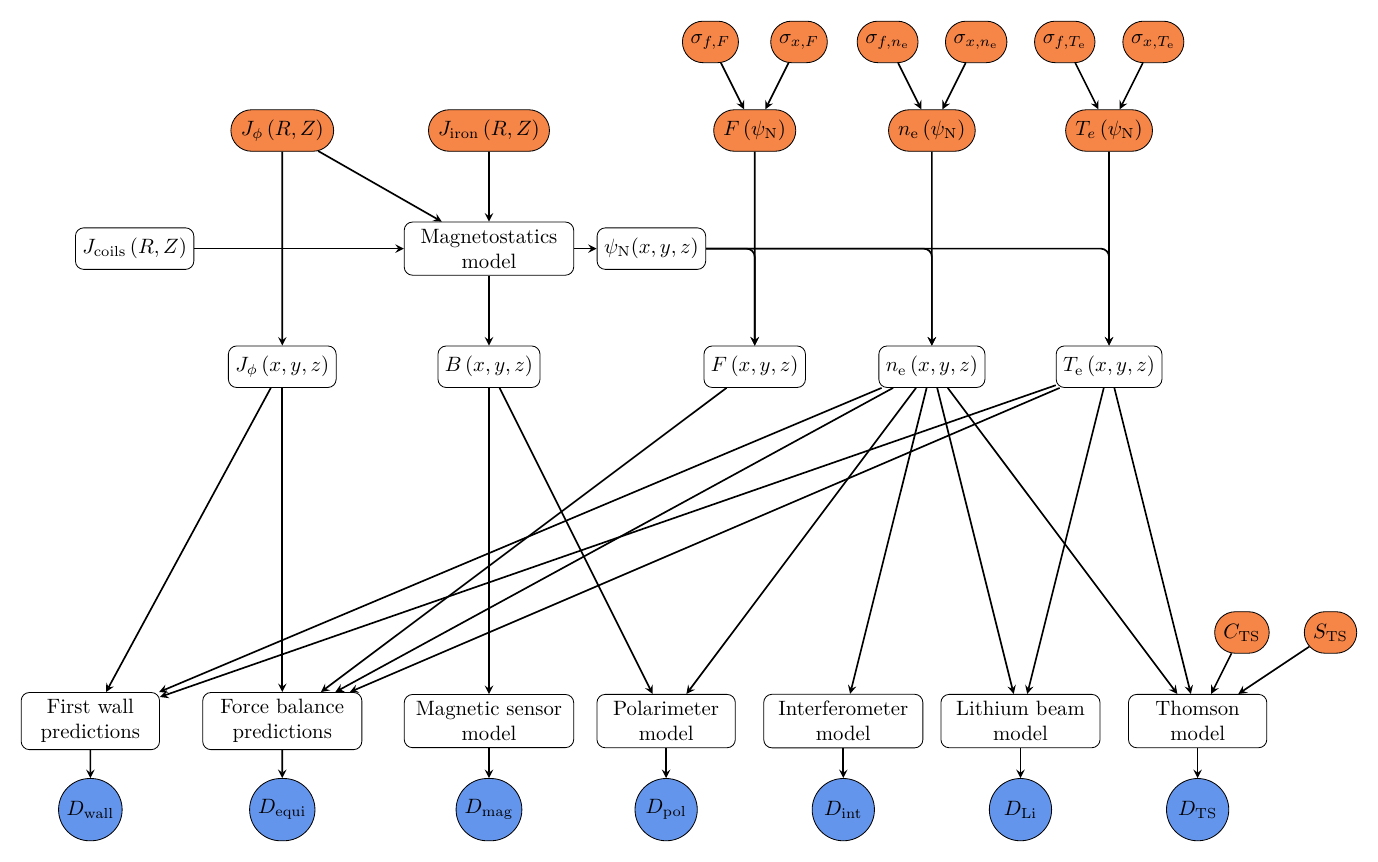}
	\caption{A simplified version of the Minerva graph representing the axisymmetric plasma equilibrium model implemented for the Joint European Torus (JET) tokamak experiment. The unknown parameters and observations are shown as red and blue circles, respectively. The toroidal plasma current $J_\phi$, iron core $J_\mathrm{iron}$ and magnetic field coils $J_\mathrm{coils}$ are modelled as a set of current-carrying solid beams. These toroidal currents determine the normalised poloidal magnetic flux $\psi_\mathrm{N}$. Plasma electron density $n_\mathrm{e}$ and temperature $T_\mathrm{e}$ and poloidal current flux $F$ are given as a function of $\psi_\mathrm{N}$. These profiles are modelled as Gaussian processes whose smoothness parameters (hyperparameters) are denoted as $\sigma_f$ and $\sigma_x$. All these physical quantities are mapped to the $x,y,z$ Cartesian coordinates. Given these quantities in real space, predictions calculated by the predictive (forward) models are compared to the observations ($D_\mathrm{mag}$, $D_\mathrm{pol}$, $D_\mathrm{int}$, $D_\mathrm{Li}$ and $D_\mathrm{TS}$). To find equilibrium solutions, we implement an MHD force balance constraint by introducing virtual observations $D_\mathrm{equi}$. In the same way, we also implement an empirical constraint at the last material surface facing the plasma inside the machine, known as the first wall (denoted as $D_\mathrm{wall}$).}
	\label{fig:model}
\end{figure}

This method involves multiple diagnostics, assumptions, unknown parameters and observations therefore it is inevitable to use a framework that is capable of handling and keeping track of them. For this reason, this method is implemented in the Minerva framework, which is developed for general scientific modelling for a complex system. This framework provides a standardised format of model components such as forward models and probability distributions and a standardised interface between these components. Minerva automatically manages all the model components and their connections which can be represented as a graphical model, as shown in Figure~\ref{fig:model}. The modular structure and automatic model administration allow us to handle a complex model systematically. Furthermore, we can easily extend Minerva models and transfer them to other experiments. In nuclear fusion research, the Minerva framework is used for a number of scientific applications for current tomography \cite{Svensson2008}, interferometer \cite{Ford2010,Svensson2011_GP}, soft X-ray \cite{Li2013, Schilling2021}, polarimeters \cite{Ford2008}, Thomson scattering \cite{Bozhenkov2017, Kwak2020}, beam emission spectroscopy \cite{Kwak2016, Kwak2017}, X-ray imaging crystal spectroscopy \cite{Langenberg2016}, electron cyclotron emission \cite{Hoefel2019} and effective ion charge \cite{Pavone2019_Zeff, Kwak2021}. These Minerva models can be accelerated by a field-programmable gate array (FPGA) \cite{Mora2017} or an artificial neural network \cite{Pavone2018, Pavone2019}.

\section{The model}\label{sec:model}
In Bayesian inference \cite{Pearl1988, Jaynes2003, Sivia2006}, a model, which embodies the full relationship between unknown parameters and observations, can be represented as a joint probability distribution $P\left(H,D\right)$. This joint distribution can be written as:
\begin{equation}
P\left(H,D\right)=P\left(D|H\right)P\left(H\right).
\label{eq:joint_distribution}
\end{equation}
The prior distribution $P\left(H\right)$ encodes model assumptions based on the \textit{prior knowledge}, for example, density or temperature must be positive. Given a hypothetical value of the unknown parameters, we can make a prediction as a \textit{predictive distribution} $P\left(D|H\right)$ over the observations. Typically, the mean of predictive distribution can be given as a function, which encapsulates the underlying processes happening during an experiment by taking into account physics as well as experimental setup, also known as a \textit{forward model} $f\left(H\right)$. The prior distribution can be updated to the posterior distribution $P\left(H|D\right)$ by the Bayes' theorem:
\begin{equation}
P\left(H|D\right)=\frac{P\left(H,D\right)}{P\left(D\right)}=\frac{P\left(D|H\right)P\left(H\right)}{P\left(D\right)},
\label{eq:Bayes_theorem}
\end{equation}
where $P\left(D\right)$ is so-called model evidence, which is a normalisation constant in this context. 

If the model contains multiple parameters and observations, the joint distribution can be calculated as a product of individual prior and predictive distributions, conditional on their parent variables:
\begin{equation}
P\left(\right\{H_i\left\},\right\{D_j\left\}\right)=\big(\prod_{j}P\left(D_j|H\right)\big)\big(\prod_{i}P\left(H_i\right)\big).
\label{eq:Bayes_model}
\end{equation}
Each of the predictive distributions contains a forward model for corresponding experimental data that might include additional unknown parameters such as calibration factors. These predictive distributions together with the prior distributions, which encode the model assumptions for the parameters, constitute the model. The model can be represented as a graphical model, which is a transparent way of unfolding its complexity.

In this work, we developed a Bayesian model for axisymmetric plasma equilibria. This model involves multiple unknown parameters, model assumptions, predictive (forward) models and observations, as shown in Figure~\ref{fig:model}. The unknown parameters (red circles) and observations (blue circles) are connected to and/or from the forward models (white boxes). The arrows visualise the dependencies between them. The model contains the following components: the axisymmetric current-carrying beam model (\texttt{Magnetostatics model}), Gaussian processes for plasma electron density $n_\mathrm{e}$, temperature $T_\mathrm{e}$ and poloidal current flux $F$, an MHD force balance constraint (\texttt{Force balance predictions}), an empirical constraint at the last material surface facing the plasma inside the machine, known as the first wall (\texttt{First wall predictions}) and forward models for plasma diagnostics, for example, Thomson scattering (\texttt{Thomson model}). These components are described in the next sections.

\subsection{Axisymmetric current-carrying beam model}
A magnetic confinement fusion device confines a fusion plasma by a magnetic field in which the plasma experience a magnetic pressure. This magnetic pressure balances out the plasma pressure and keeps the plasma in an MHD equilibrium state. To model this MHD equilibrium, we have to model the electric currents in the plasma and the external coils. These electric currents can be separated into toroidal and poloidal parts in an axisymmetric device like a tokamak.

\begin{figure}
	\centering
	\includegraphics[width=\linewidth]{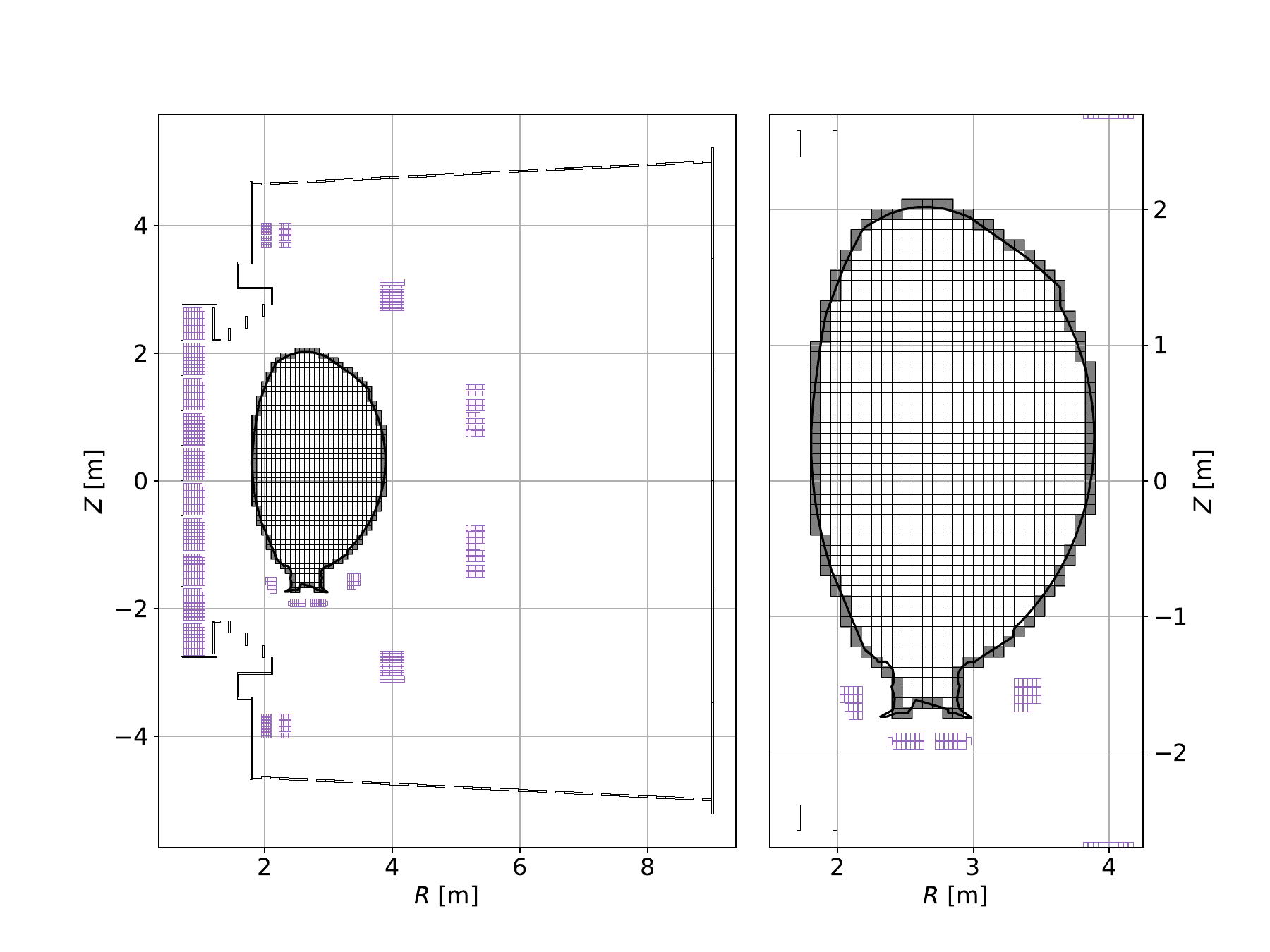}
	\caption{The axisymmetric current-carrying beam model for the JET tokamak. The beam grids for toroidal plasma current $J_\phi$ and iron core current $J_\mathrm{iron}$ (unknown) are in black. On the other hand, the beam grid for the poloidal magnetic field coils $J_\mathrm{coils}$ (known) is in purple. The current-carrying beams on the first wall (thick black line), are filled with grey.}
	\label{fig:current_model}
\end{figure}

In this work, we use the axisymmetric current-carrying beam model, which is previously developed for current tomography \cite{Svensson2008}. In this model, the toroidal electric currents are modelled as a set of axisymmetric current-carrying solid beams with finite rectangular cross-sections (Figure~\ref{fig:current_model}). We take into account toroidal plasma current $J_\phi$, iron core $J_\mathrm{iron}$ and magnetic field coils $J_\mathrm{coils}$ at the JET tokamak experiment. $J_\phi$ and $J_\mathrm{iron}$ are unknown parameters in the model, on the other hand, $J_\mathrm{coils}$ is recorded in the JET database during an experiment (known parameter).

Given a toroidal current density $J$, the magnetic vector potential $A$ at a spatial location $r=\left[x,y,z\right]$ can be calculated by the Biot-Savart law:
\begin{equation}
A\left(r\right)=\frac{\mu_0}{4\pi}\iiint\frac{J\left(r^\prime\right)}{\left\lvert r-r^\prime\right\rvert}\,\mathrm{d}^{3}r^\prime,
\label{eq:Biot-Savart}
\end{equation}
where $\mu_0$ is the vacuum permeability. This calculation can be done in a short time by multiplying the current density by a pre-calculated response factor. We calculated this response factor for every location where we have to determine the magnetic field, for instance, magnetic probe positions. The magnetic field $B$ can be calculated from the vector potential:
\begin{equation}
B=\nabla\times A.
\label{eq:magentic_field}
\end{equation}

To obtain the magnetic field geometry of the plasma, we calculate poloidal magnetic flux:
\begin{equation}
\psi\left(x,y,z\right)=\oint A\cdot\,\mathrm{d}\ell.
\end{equation}
This poloidal magnetic flux can be normalised to zero at the magnetic axis and to one at the LCFS. This normalised flux $\psi_\mathrm{N}$ is often used as the canonical coordinate system for further data analysis and physics studies. We map the other physical quantities such as $n_\mathrm{e}$ and $T_\mathrm{e}$ to $\psi_\mathrm{N}$ and model them as non-parametric Gaussian processes.

\subsection{Gaussian process prior}
A Gaussian process \cite{OHagan1978, Neal1995, Rasmussen2006} is a non-parametric function that associates a set of input points on the domain (e.g. space and time) with a set of output values, which follows a multivariate Gaussian distribution. In this context, this function can be seen as a (long) vector, containing an output value $f\left(x\right)$ at every input point $x$. As one might imagine, even though we do not define any parametric formula for the function, the output values along the input points are normally distributed with a certain correlation, which determines the smoothness of the function. This correlation between any two points is defined by another function, known as the covariance function. On the other hand, the mean function, which gives an output mean at every input point, determines the general \textit{prior} trend of the output distribution. Since the Gaussian process is a function defined by a Gaussian distribution, it can be seen as a generalisation of the Gaussian distribution to a function space. Unlike a parametric model, which typically restricts a solution in a specific shape such as a parabola, a Gaussian process does not have any particular parameterisation for the function output. Instead, we can determine the behaviour of the process, for instance, smoothness or periodicity by the mean and covariance function. In nuclear fusion research, Gaussian processes were introduced by a non-parametric tomography method for electron density and plasma current distribution \cite{Svensson2011_GP}, followed by several applications \cite{Chilenski2015, Kwak2016, Langenberg2016, Kwak2017, Romero2018, Kwak2020}.

One of the most widely used mean and covariance functions in Gaussian processes is a zero mean function and a squared exponential covariance function. A Gaussian process $f$ with these mean and covariance functions can be written as:
\begin{align}
f\left(x\right)&\sim\mathcal{N}\left(\mu\left(x\right),\Sigma\left(x,x\right)\right),\label{eq:Gaussian_process}\\
\mu\left(x\right)&=0,\label{eq:zero_mean}\\
\Sigma\left(x_i,x_j\right)&=\sigma_f^2\exp{\left(-\frac{\left(x_i-x_j\right)^2}{2\sigma_x^2}\right)}+\sigma_y^2\delta_{ij}.\label{eq:squared_exponential_covariance}
\end{align}
Here, $\mu$ is a zero mean function, implying that the function has no trend a~priori. The covariance function $\Sigma\left(x_i,x_j\right)$ gives the covariance value between two arbitrary points $x_i$ and $x_j$. The overall scale $\sigma_f$ and the length scale $\sigma_x$, which are so-called hyperparameters, determine the smoothness of the function. $\sigma_y$ is chosen to be a relatively small number, for example, $\sigma_y/\sigma_f=10^{-3}$ to avoid numerical instabilities.

The prior distribution of poloidal current flux $F$ can be modelled as this Gaussian process:
\begin{equation}
P\left(F|\sigma_{f,F},\sigma_{x,F}\right)=\mathcal{N}\left(\mu_F\left(\psi_\mathrm{N}\right),\Sigma_F\left(\psi_\mathrm{N},\psi_\mathrm{N}\right)\right),
\label{eq:prior_F}
\end{equation}
where $\mu_F$ and $\Sigma_F$ are given by Equation~(\ref{eq:zero_mean}) and Equation~(\ref{eq:squared_exponential_covariance}), respectively. The prior distributions of the hyperparameters $\sigma_F=\left[\sigma_{f,F},\sigma_{x,F}\right]$ are given as an uniform distribution.

Electron density $n_\mathrm{e}$ and temperature $T_\mathrm{e}$ might have substantially different gradient (smoothness) in the core and edge regions \cite{ASDEXTeam1989}. In this case, we use a non-stationary covariance function \cite{Higdon1999} for spatially varying smoothness:
\begin{equation}
\Sigma\left(x_i,x_j\right)=\sigma_f^2\left(\frac{2\sigma_x\left(x_i\right)\sigma_x\left(x_j\right)}{\sigma_x\left(x_i\right)^2+\sigma_x\left(x_j\right)^2}\right)^{\frac{1}{2}}\exp{\left(-\frac{\left(x_i-x_j\right)^2}{\sigma_x\left(x_i\right)^2+\sigma_x\left(x_j\right)^2}\right)}+\sigma_y^2\delta_{ij},
\label{eq:Higdon_covariance}
\end{equation}
where the length scale $\sigma_x\left(x\right)$ can be given as an arbitrary function. Here, we choose a hyperbolic tangent function for a smooth transition between the core and edge gradient values \cite{Chilenski2015, Kwak2020}:
\begin{equation}
\sigma_x\left(x\right)=\frac{\sigma_{x,\mathrm{core}}+\sigma_{x,\mathrm{edge}}}{2}-\frac{\sigma_{x,\mathrm{core}}-\sigma_{x,\mathrm{edge}}}{2}\tanh{\left(\frac{x-x_0}{x_\mathrm{w}}\right)},
\label{eq:length_scale_function}
\end{equation}
where $\sigma_{x,\mathrm{core}}$ and $\sigma_{x,\mathrm{edge}}$ are the length scales in the core and edge regions. The position and width of the smoothness (gradient) transition are denoted as $x_0$ and $x_\mathrm{w}$. The prior distributions of $n_\mathrm{e}$ and $T_\mathrm{e}$ can be modelled as this Gaussian process:
\begin{align}
P\left(n_\mathrm{e}|\sigma_{f,n_\mathrm{e}},\sigma_{x,n_\mathrm{e}}\right)=\mathcal{N}\left(\mu_{n_\mathrm{e}}\left(\psi_\mathrm{N}\right),\Sigma_{n_\mathrm{e}}\left(\psi_\mathrm{N},\psi_\mathrm{N}\right)\right),\label{eq:prior_ne}\\
P\left(T_\mathrm{e}|\sigma_{f,T_\mathrm{e}},\sigma_{x,T_\mathrm{e}}\right)=\mathcal{N}\left(\mu_{T_\mathrm{e}}\left(\psi_\mathrm{N}\right),\Sigma_{T_\mathrm{e}}\left(\psi_\mathrm{N},\psi_\mathrm{N}\right)\right),\label{eq:prior_Te}
\end{align}
where $\mu_{n_\mathrm{e}}$, $\mu_{T_\mathrm{e}}$, $\Sigma_{n_\mathrm{e}}$ and $\Sigma_{T_\mathrm{e}}$ are given by Equation~(\ref{eq:zero_mean}) and Equation~(\ref{eq:Higdon_covariance}). Each of the length scales $\sigma_{x,n_\mathrm{e}}$ and $\sigma_{x,T_\mathrm{e}}$ is given by Equation~(\ref{eq:length_scale_function}) and contains the four hyperparameters $\sigma_{x,\mathrm{core}}$, $\sigma_{x,\mathrm{edge}}$, $x_0$ and $x_\mathrm{w}$, e.g., $\sigma_{x,n_\mathrm{e}}=\left[\sigma_{x,n_\mathrm{e},\mathrm{core}},\sigma_{x,n_\mathrm{e},\mathrm{edge}},x_{0,n_\mathrm{e}},x_\mathrm{w,n_\mathrm{e}}\right]$. Again, the prior distributions of these hyperparameters $\sigma_{n_\mathrm{e}}=\left[\sigma_{f,n_\mathrm{e}},\sigma_{x,n_\mathrm{e}}\right]$ and $\sigma_{T_\mathrm{e}}=\left[\sigma_{f,T_\mathrm{e}},\sigma_{x,T_\mathrm{e}}\right]$ are given as an uniform distribution.

All these physical quantities modelled as 1D Gaussian processes can be mapped to real space. Given all these quantities as 3D fields, we can calculate predictions for experimental data or derived quantities in physics equations. This means that we can examine not only experimental data but also physics equations. For instance, we can compute both sides of the Grad-Shafranov MHD force balance equation and compare them at any spatial location. Furthermore, by imposing that the right- and left-hand side sides of the force balance equation must be equal, we can introduce an MHD force balance constraint.

\subsection{The equilibrium prior}
As described previously, a plasma can be confined in an equilibrium state in which the plasma pressure gradient is balanced out by the magnetic force. This MHD equilibrium can be described by the MHD force balance equation:
\begin{equation}
J\times B-\nabla p\simeq 0,
\label{eq:MHD_force_balance}
\end{equation}
where $J$ is the plasma current density, $B$ the magnetic field and $p$ the isotropic plasma pressure. For an axisymmetric plasma, this force balance can be given in terms of toroidal current density $J_\phi$, poloidal current flux $F$ and pressure $p$ by the Grad-Shafranov equation \cite{Grad1958, Shafranov1963}:
\begin{equation}
J_\phi-Rp^\prime-\frac{\mu_0}{R}FF^\prime\simeq 0,
\label{eq:GS_force_balance}
\end{equation}
where $p^\prime=\frac{\partial p}{\partial\psi}$ and $F^\prime=\frac{\partial F}{\partial\psi}$. To examine the MHD force balance for a plasma current beam, we can integrate this equation over the beam cross-section. This MHD force balance constraint can be implemented by introducing \textit{virtual observations} \cite{Ford2010}, which can be written as:
\begin{equation}
P\left(D_\mathrm{equi}|J_\phi,n_\mathrm{e},T_\mathrm{e},F\right)=\prod_{i}\mathcal{N}\left(\int_{Z_{\mathrm{min},i}}^{Z_{\mathrm{max},i}}\int_{R_{\mathrm{min},i}}^{R_{\mathrm{max},i}}J_\phi-Rp^\prime-\frac{\mu_0}{R}FF^\prime\,\mathrm{d}R\,\mathrm{d}Z,\sigma_\mathrm{equi}\right),
\label{eq:equilibrium_vobs}
\end{equation}
where $R_{\mathrm{min},i}$, $R_{\mathrm{max},i}$, $Z_{\mathrm{min},i}$ and $Z_{\mathrm{max},i}$ define the rectangular cross-section of the $i^\mathrm{th}$ plasma beam. For plasma pressure, we assume $p=2n_\mathrm{e}T_\mathrm{e}$. The \textit{observed} data $D_\mathrm{equi}$ are set to be zero. This implies that the Grad-Shafranov equation should be fulfilled. The uncertainties of the virtual observations is set to be \SI{50}{\kilo\ampere\per\metre\squared} which is a few per cent of a typical average plasma current density at JET (\SI{\approx e3}{\kilo\ampere\per\metre\squared}). These virtual observations together with the prior distributions of $J_\phi$, $n_\mathrm{e}$, $T_\mathrm{e}$ and $F$ constitute the equilibrium prior:
\begin{equation}
P\left(J_\phi,n_\mathrm{e},T_\mathrm{e},F|D_\mathrm{equi}\right)=\frac{P\left(D_\mathrm{equi}|J_\phi,n_\mathrm{e},T_\mathrm{e},F\right)P\left(J_\phi\right)P\left(n_\mathrm{e}\right)P\left(T_\mathrm{e}\right)P\left(F\right)}{P\left(D_\mathrm{equi}\right)},
\label{eq:equilibrium_prior}
\end{equation}
where $P\left(J_\phi\right)$ is chosen to be a Gaussian distribution with a zero mean and a standard deviation of \SI{300e6}{\kilo\ampere\per\metre\squared}, which is effectively an uniform for $J_\phi$ at JET.

For comparison, we also calculate solutions without the equilibrium prior. In this case, the prior distribution of $J_\phi$ is modelled as a Gaussian process with a zero mean function and a squared exponential covariance function:
\begin{equation}
P\left(J_\phi|\sigma_{f,J_\phi},\sigma_{x,J_\phi}\right)=\mathcal{N}\left(\mu_{J_\phi}\left(x\right),\Sigma_{J_\phi}\left(x,x\right)\right),
\label{eq:prior_J}
\end{equation}
where $\mu_{J_\phi}$ and $\Sigma_{J_\phi}$ are given by Equation~(\ref{eq:zero_mean}) and Equation~(\ref{eq:squared_exponential_covariance}), respectively. We note that $\mu_{J_\phi}$ and $\Sigma_{J_\phi}$ are a function of $x=\left[R,Z\right]$. This means that we have the length scales for $R$ and $Z$, i.e., $\sigma_{x,J_\phi}=\left[\sigma_{R,J_\phi},\sigma_{Z,J_\phi}\right]$. The prior distributions of the hyperparameters $\sigma_{J_\phi}=\left[\sigma_{f,J_\phi},\sigma_{x,J_\phi}\right]$ are given as an uniform distribution.

\subsection{Plasma diagnostics}
\begin{figure}
	\centering
	\includegraphics[width=\linewidth]{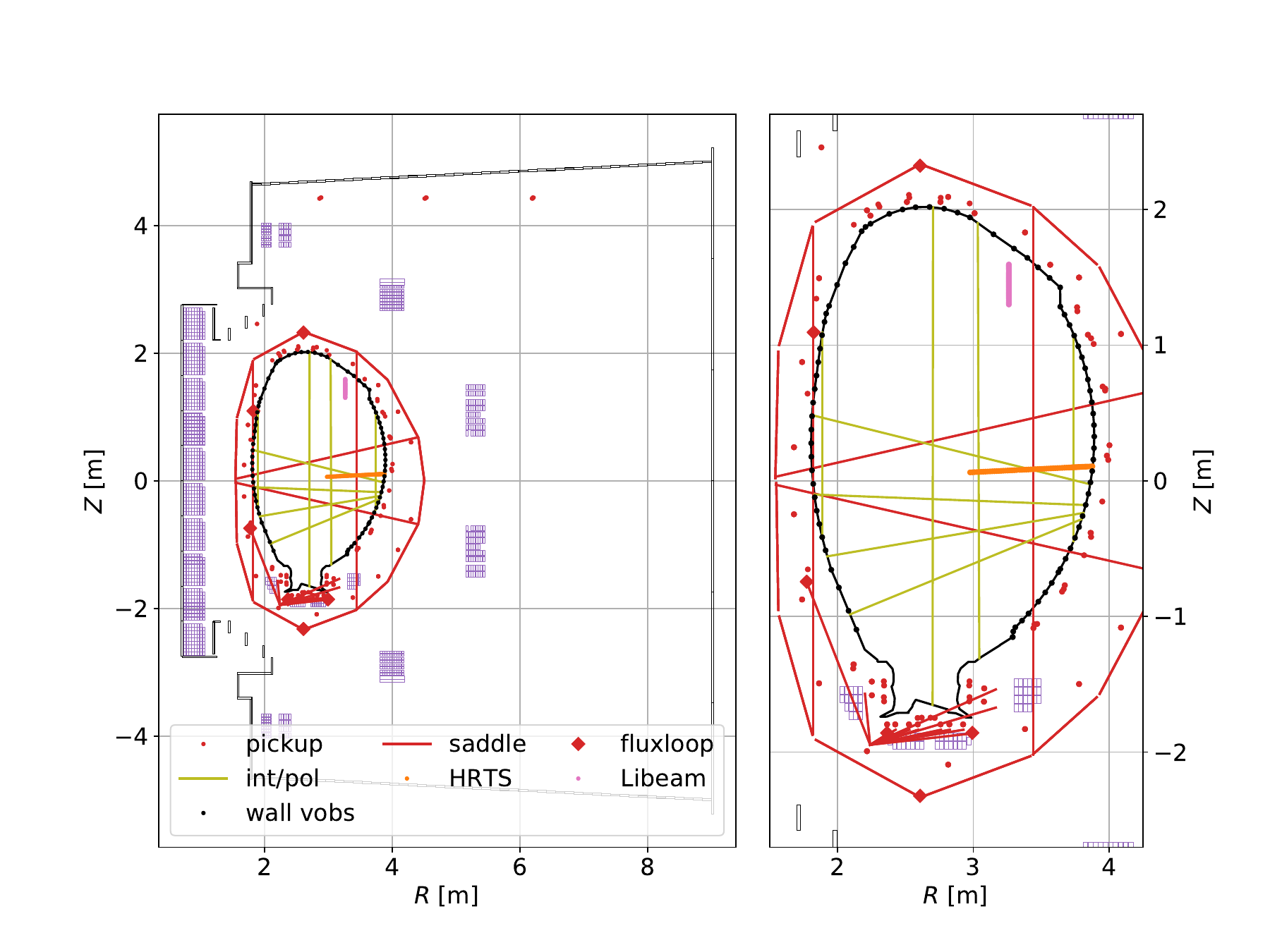}
	\caption{The measurement positions and lines of sight of the diagnostics used in this model, projected into a single poloidal plane. The magnetic probes including pickup coils, saddle coils and flux loops (in red) measure the magnetic field. The interferometers and polarimeters share the same lines of sight (in yellow) and provide measurements for the following line-integrated quantities: $\int n_\mathrm{e}\,\mathrm{d}\ell$ and $\int n_\mathrm{e}B_\parallel\,\mathrm{d}\ell$. The high-resolution Thomson scattering (HRTS) system measures $n_\mathrm{e}$ and $T_\mathrm{e}$ at $63$ spatial locations (in orange) along a laser path near the mid-plane. The lithium beam emission spectroscopy system provides edge $n_\mathrm{e}$ at $26$ spatial locations (in pink) along the vertically injected lithium beam from the top. The first wall and the positions of the wall constraint are shown as a black line and black dots, respectively.}
	\label{fig:diagnsotics}
\end{figure}

We modelled plasma current and pressure distributions and implemented the equilibrium constraint in the model. To infer equilibrium current and pressure distributions, we should take into account magnetic field and plasma pressure measurements. In this work, we employ multiple plasma diagnostics: magnetic probes (pickup coils, saddle coils and flux loops), polarimeters, interferometers, high-resolution Thomson scattering (HRTS) and lithium beam emission spectroscopy systems (Figure~\ref{fig:diagnsotics}). We use the forward models for these diagnostics, which are previously developed in other applications \cite{Svensson2008, Ford2008, Svensson2011_GP, Kwak2016, Kwak2017, Kwak2020}, for this work with several improvements. These forward models thoroughly encapsulate all the relevant physics and experimental setup, which are briefly described in the following subsections.

\subsubsection{Magnetic probes}
To take into account magnetic field measurements, we use the magnetic probes including pickup coils, saddle coils and full flux loops at JET (red dots, lines and diamonds in Figure~\ref{fig:diagnsotics}). A pickup coil measures the local magnetic field at its position (red dot). A saddle coil covering one of the octants between two poloidal positions (shown as a red line defined by these two positions) measures magnetic flux through it. A full flux loop provides total magnetic flux through a circular surface defined by a poloidal position (red diamond). The predictive model for all these magnetic probes can be written as:
\begin{align}
P\left(D_\mathrm{mag}|J_\phi,J_\mathrm{iron}\right)&=P\left(D_\mathrm{pickup}|J_\phi,J_\mathrm{iron}\right)P\left(D_\mathrm{saddle}|J_\phi,J_\mathrm{iron}\right)P\left(D_\mathrm{fluxloop}|J_\phi,J_\mathrm{iron}\right),\label{eq:predictive_magnetic_sensors}\\
P\left(D_\mathrm{pickup}|J_\phi,J_\mathrm{iron}\right)&=\prod_{i}\mathcal{N}\left(B_R\left(R_i,Z_i\right)\cos{\theta_i}+B_Z\left(R_i,Z_i\right)\sin{\theta_i},\sigma_{\mathrm{pickup},i}\right),\label{eq:predictive_pickups}\\
P\left(D_\mathrm{saddle}|J_\phi,J_\mathrm{iron}\right)&=\prod_{i}\mathcal{N}\left(\frac{G_{\mathrm{saddle},i}}{8}\left(\psi\left(R_{2,i},Z_{2,i}\right)-\psi\left(R_{1,i},Z_{1,i}\right)\right),\sigma_{\mathrm{saddle},i}\right),\label{eq:predictive_saddles}\\
P\left(D_\mathrm{fluxloop}|J_\phi,J_\mathrm{iron}\right)&=\prod_{i}\mathcal{N}\left(\psi\left(R_i,Z_i\right),\sigma_{\mathrm{fluxloop},i}\right),\label{eq:predictive_fluxloops}
\end{align}
where $\theta_i$ is the angle of the normal vector of the $i^\mathrm{th}$ pickup coil, $\left(R_{1,i},Z_{1,i}\right)$ and $\left(R_{2,i},Z_{2,i}\right)$ are the two poloidal positions of the $i^\mathrm{th}$ saddle coil and $G_{\mathrm{saddle},i}$ the geometry factor for taking into account the 3D geometry of the $i^\mathrm{th}$ saddle coil. The magnetic probe data $D_\mathrm{mag}=\left[D_\mathrm{pickup},D_\mathrm{saddle},D_\mathrm{fluxloop}\right]$ and their uncertainties $\sigma_\mathrm{mag}=\left[\sigma_\mathrm{pickup},\sigma_\mathrm{saddle},\sigma_\mathrm{fluxloop}\right]$ are retrieved from the JET database.

%where $B_R\left(R_i,Z_i\right)$ and $B_Z\left(R_i,Z_i\right)$ are the $R$ and $Z$ direction of the magnetic field at the position of the $i^\mathrm{th}$ pickup coil with the angle of the normal vector $\theta_i$, $\psi\left(R_{2,i},Z_{2,i}\right)$ and $\psi\left(R_{1,i},Z_{1,i}\right)$ are the poloidal magnetic flux at the two $R,Z$ positions of the $i^\mathrm{th}$ saddle coil with the correction factor for the actual 3D coil geometry $G_{\mathrm{saddle},i}$ and $\psi\left(R_i,Z_i\right)$ is the poloidal magnetic flux through the $i^\mathrm{th}$ flux loop. All the observations and uncertainties of all these magnetic sensors $D_\mathrm{mag}=\left[D_\mathrm{pickup},D_\mathrm{saddle},D_\mathrm{fluxloop}\right]$ and $\sigma_\mathrm{mag}=\left[\sigma_\mathrm{pickup},\sigma_\mathrm{saddle},\sigma_\mathrm{fluxloop}\right]$ are retrieved from the JET database.

The 3D fields of $n_\mathrm{e}$, $T_\mathrm{e}$ and $F$ are mapped from their 1D Gaussian processes profiles on the $\psi_\mathrm{N}$ coordinates. Thus, $n_\mathrm{e}$, $T_\mathrm{e}$ and $F$ profiles have a \textit{hidden} dependence on $J_\phi$ and $J_\mathrm{iron}$ and $J_\mathrm{coils}$, but this will be omitted to avoid confusion.

\subsubsection{Interferometers and polarimeters}
The far-infrared (FIR) interferometer-polarimeter system at JET \cite{Boboc2010, Boboc2012, Boboc2015} launches electromagnetic waves into the plasma and measures the phase differences and the Faraday rotations between the injected and reference waves which are proportional to the following line integrated quantities: $\int n_\mathrm{e}\,\mathrm{d}\ell$ and $\int n_\mathrm{e}B_\parallel\,\mathrm{d}\ell$. The system has four lateral and four vertical lines of sight, as shown in Figure~\ref{fig:diagnsotics}. The predictive model for the interferometer-polarimeter system can be written as:
\begin{align}
P\left(D_\mathrm{int}|n_\mathrm{e}\left(\psi_\mathrm{N}\right)\right)&=\prod_{i}\mathcal{N}\left(\int n_\mathrm{e}\,\mathrm{d}\ell_i,\sigma_{\mathrm{int},i}\right),\label{eq:predictive_interferometers}\\
P\left(D_\mathrm{pol}|J_\phi,n_\mathrm{e}\left(\psi_\mathrm{N}\right)\right)&=\prod_{i}\mathcal{N}\left(\int n_\mathrm{e}B_\parallel\,\mathrm{d}\ell_i,\sigma_{\mathrm{pol},i}\right),\label{eq:predictive_polarimeters}
\end{align}
where $\int\,\mathrm{d}\ell_i$ is a line integral along the $i^\mathrm{th}$ line of sight and $B_\parallel$ the magnetic field strength parallel to the line of sight. The line integrated data $D_\mathrm{int}$ and $D_\mathrm{pol}$ and their uncertainties $\sigma_\mathrm{int}$ and $\sigma_\mathrm{pol}$ are retrieved from the JET database.

\subsubsection{High-resolution Thomson scattering system}
The high-resolution Thomson scattering (HRTS) system at JET \cite{Pasqualotto2004} launches laser pulses into the plasma and collects Thomson scattered spectra \cite{Thomson1906} by polychromators with four spectral channels from $63$ spatial locations (orange dots in Figure~\ref{fig:diagnsotics}). This system has a spatial resolution of \SIrange{0.8}{1.6}{\centi\metre} and a temporal resolution of \SI{20}{\hertz}. The intensity and width of Thomson scattering spectra provide $n_\mathrm{e}$ and $T_\mathrm{e}$ measurements. The $n_\mathrm{e}$ calibration factor $C_\mathrm{TS}$ and the position shift $S_\mathrm{TS}$ of all spatial channels along the laser path are regarded as additional unknown parameters in this model. The predictive model for the HRTS system can be written as:
\begin{align}
&P\left(D_\mathrm{TS}|n_\mathrm{e}\left(\psi_\mathrm{N}\right),T_\mathrm{e}\left(\psi_\mathrm{N}\right),C_\mathrm{TS},S_\mathrm{TS}\right)\nonumber\\
=&\prod_{i}\prod_{j}\mathcal{N}\left(A_{\mathrm{TS},i,j}\left(n_\mathrm{e}\left(R_i,Z_i,S_\mathrm{TS}\right),T_\mathrm{e}\left(R_i,Z_i,S_\mathrm{TS}\right),C_\mathrm{TS}\right),\sigma_{\mathrm{TS},i,j}\right),\label{eq:predictive_HRTS}
\end{align}
where $A_{\mathrm{TS},i,j}$ is the amplitude of the Thomson scattering spectrum of the $j^\mathrm{th}$ spectral channel of the $i^\mathrm{th}$ spatial position and $\sigma_{\mathrm{TS},i,j}$ the corresponding uncertainties. The spatially shifted $n_\mathrm{e}$ and $T_\mathrm{e}$ can be calculated as:
\begin{align}
n_\mathrm{e}\left(R_i,Z_i,S_\mathrm{TS}\right)&=n_\mathrm{e}\left(R_i+S_\mathrm{TS}\cos{\theta_\mathrm{TS}},Z_i+S_\mathrm{TS}\sin{\theta_\mathrm{TS}}\right),\label{eq:TS_denisty}\\
T_\mathrm{e}\left(R_i,Z_i,S_\mathrm{TS}\right)&=T_\mathrm{e}\left(R_i+S_\mathrm{TS}\cos{\theta_\mathrm{TS}},Z_i+S_\mathrm{TS}\sin{\theta_\mathrm{TS}}\right),\label{eq:TS_temperature}
\end{align}
where $\theta_\mathrm{TS}$ is the angle of the laser path. If $S_\mathrm{TS}$ is positive, the shift would be outward (closer to the first wall), otherwise inward. We only allow $S_\mathrm{TS}$ to be all the shifted spatial positions inside the first wall. The amplitude of the Thomson scattering spectrum can be written as:
%where $R_i$ and $Z_i$ are the position of the $i^\mathrm{th}$ spatial channel and $\theta_\mathrm{TS}$ is the angle of the laser path. If the position shift $S_\mathrm{TS}$ is positive, the shifted positions will be closer to the first wall than the original positions. The range of $S_\mathrm{TS}$ is set not to allow any shifted position to be beyond the first wall. The amplitude of the Thomson scattering spectrum can be written as:
\begin{equation}
A_{\mathrm{TS},i,j}\left(n_\mathrm{e},T_\mathrm{e},C_\mathrm{TS}\right)=C_\mathrm{TS}\,n_\mathrm{e}E_\mathrm{laser}\int\phi_{i,j}\left(\lambda\right)\frac{\lambda}{hc}r_\mathrm{e}^2\frac{S\left(\lambda,\theta,T_\mathrm{e}\right)}{\lambda_\mathrm{laser}}\,\mathrm{d}\lambda,\label{eq:TS_model}
\end{equation}
where $E_\mathrm{laser}$ is the laser energy, $\phi_{i,j}\left(\lambda\right)$ spectral response function of the $j^\mathrm{th}$ spectral channel of the $i^\mathrm{th}$ spatial position, $\lambda$ the scattered wavelength, $h$ the Planck constant, $c$ the speed of light, $r_\mathrm{e}$ the classical electron radius, $S\left(\lambda,\theta,T_\mathrm{e}\right)$ the spectral density function \cite{Naito1993}, $\theta$ the scattering angle and $\lambda_\mathrm{laser}$ the laser wavelength. The prior distributions of $C_\mathrm{TS}$ and $S_\mathrm{TS}$ are given as an uniform distribution.

\subsubsection{Lithium beam emission spectroscopy system}
The lithium beam emission spectroscopy system at JET \cite{Brix2010, Brix2012} injects lithium beam atoms into the plasma and collects line emission at $26$ spatial locations (pink dots in Figure~\ref{fig:diagnsotics}) with a spatial resolution of \SI{\approx1.0}{\centi\metre} and a temporal resolution of \SIrange{10}{20}{\milli\second}. Some of the lithium atoms can be raised to the first excited state by electron- and ion-impact excitation, and these excited atoms may produce the line emission spontaneously. The intensity of the lithium line emission can be used to infer both $n_\mathrm{e}$ and $T_\mathrm{e}$ in principle, but in practice, it is normally used to get only $n_\mathrm{e}$. The lithium beam system at JET is designed to provide edge $n_\mathrm{e}$ profiles. The predictive model for the lithium beam system can be written as:
\begin{equation}
P\left(D_\mathrm{Li}|n_\mathrm{e}\left(\psi_\mathrm{N}\right),T_\mathrm{e}\left(\psi_\mathrm{N}\right)\right)=\prod_{i}\mathcal{N}\left(A_{\mathrm{Li},i}\left(n_\mathrm{e}\left(x_i,y_i,z_i\right),T_\mathrm{e}\left(x_i,y_i,z_i\right)\right),\sigma_{\mathrm{Li},i}\right),\label{eq:predictive_Li}
\end{equation}
where $A_{\mathrm{Li},i}$ is the lithium line emission intensity of the $i^\mathrm{th}$ spatial position. The line emission intensity can be calculated by the collisional-radiative model, which takes into account excitation and de-excitation, ionisation and spontaneous emission \cite{Kwak2017}. The data $D_\mathrm{Li}$ and their uncertainties $\sigma_\mathrm{Li}$ are retrieved from the JET database.

\subsection{The wall constraint}
During an experiment, $J_\phi$, $n_\mathrm{e}$ and $T_\mathrm{e}$ should not be too high on the last material surface facing the plasma inside the machine, known as the first wall. This boundary condition can be implemented by introducing another set of virtual observations at the outermost plasma current beams for $J_\phi$ (shaded beams in Figure~\ref{fig:current_model}) and on the first wall except for the divertor region for $n_\mathrm{e}$ and $T_\mathrm{e}$ (black dots in Figure~\ref{fig:diagnsotics}):
\begin{align}
P\left(D_\mathrm{wall}|J_\phi,n_\mathrm{e},T_\mathrm{e}\right)&=P\left(D_{\mathrm{wall},J_\phi}|J_\phi\right)P\left(D_{\mathrm{wall},n_\mathrm{e}}|n_\mathrm{e}\right)P\left(D_{\mathrm{wall},T_\mathrm{e}}|T_\mathrm{e}\right),\label{eq:wall_vobs}\\
P\left(D_{\mathrm{wall},J_\phi}|J_\phi\right)&=\prod_{i}\mathcal{N}\left(J_\phi\left(R_i,Z_i\right),\sigma_{\mathrm{wall},J_\phi}\right),\label{eq:wall_vobs_current}\\
P\left(D_{\mathrm{wall},n_\mathrm{e}}|n_\mathrm{e}\right)&=\prod_{i}\mathcal{N}\left(n_\mathrm{e}\left(x_i,y_i,z_i\right),\sigma_{\mathrm{wall},n_\mathrm{e}}\right),\label{eq:wall_vobs_density}\\
P\left(D_{\mathrm{wall},T_\mathrm{e}}|T_\mathrm{e}\right)&=\prod_{i}\mathcal{N}\left(T_\mathrm{e}\left(x_i,y_i,z_i\right),\sigma_{\mathrm{wall},T_\mathrm{e}}\right),\label{eq:wall_vobs_temperature}
\end{align}
where $\left(R_i,Z_i\right)$ is the $i^\mathrm{th}$ outermost plasma current beam position and $\left(x_i,y_i,z_i\right)$ is the $i^\mathrm{th}$ position on the first wall. Here, we select some reasonable values for $D_\mathrm{wall}$ and $\sigma_\mathrm{wall}$: $D_{\mathrm{wall},J_\phi}=\SI{0.0}{\kilo\ampere\per\metre\squared}$, $\sigma_{\mathrm{wall},J_\phi}=\SI{1.0}{\kilo\ampere\per\metre\squared}$, $D_{\mathrm{wall},n_\mathrm{e}}=\SI{e15}{\per\metre\cubed}$, $\sigma_{\mathrm{wall},n_\mathrm{e}}=\SI{e15}{\per\metre\cubed}$, $D_{\mathrm{wall},T_\mathrm{e}}=\SI{0.1}{\electronvolt}$ and $\sigma_{\mathrm{wall},T_\mathrm{e}}=\SI{0.1}{\electronvolt}$.

\subsection{The joint distribution}
By collecting all these prior distributions and predictive models, we can construct the model as a joint distribution, which embodies the full relationship between the unknown parameters and observations. In this work, we build the model with and without the equilibrium prior. The axisymmetric plasma model without the equilibrium prior can be written as:
\begin{align}
&P\left(J_\phi,\sigma_{J_\phi},J_\mathrm{iron},n_\mathrm{e},\sigma_{n_\mathrm{e}},T_\mathrm{e},\sigma_{T_\mathrm{e}},C_\mathrm{TS},S_\mathrm{TS},D_\mathrm{mag},D_\mathrm{int},D_\mathrm{pol},D_\mathrm{TS},D_\mathrm{Li},D_\mathrm{wall}\right)\nonumber\\
=&P\left(D_\mathrm{mag}|J_\phi,J_\mathrm{iron}\right)P\left(D_\mathrm{int}|n_\mathrm{e}\left(\psi_\mathrm{N}\right)\right)P\left(D_\mathrm{pol}|J_\phi,J_\mathrm{iron},n_\mathrm{e}\left(\psi_\mathrm{N}\right)\right)\nonumber\\
\times&P\left(D_\mathrm{TS}|n_\mathrm{e}\left(\psi_\mathrm{N}\right),T_\mathrm{e}\left(\psi_\mathrm{N}\right),C_\mathrm{TS},S_\mathrm{TS}\right)P\left(C_\mathrm{TS}\right)P\left(S_\mathrm{TS}\right)P\left(D_\mathrm{Li}|n_\mathrm{e}\left(\psi_\mathrm{N}\right),T_\mathrm{e}\left(\psi_\mathrm{N}\right)\right)\nonumber\\
\times&P\left(D_\mathrm{wall}|J_\phi,n_\mathrm{e}\left(\psi_\mathrm{N}\right),T_\mathrm{e}\left(\psi_\mathrm{N}\right)\right)P\left(J_\phi|\sigma_{f,J_\phi},\sigma_{x,J_\phi}\right)P\left(\sigma_{f,J_\phi}\right)P\left(\sigma_{x,J_\phi}\right)P\left(J_\mathrm{iron}\right)\nonumber\\
\times&P\left(n_\mathrm{e}|\sigma_{f,n_\mathrm{e}},\sigma_{x,n_\mathrm{e}}\right)P\left(\sigma_{f,n_\mathrm{e}}\right)P\left(\sigma_{x,n_\mathrm{e}}\right)P\left(T_\mathrm{e}|\sigma_{f,T_\mathrm{e}},\sigma_{x,T_\mathrm{e}}\right)P\left(\sigma_{f,T_\mathrm{e}}\right)P\left(\sigma_{x,T_\mathrm{e}}\right),\label{eq:CT_joint_distribution}
\end{align}
and with the equilibrium prior:
\begin{align}
&P\left(J_\phi,J_\mathrm{iron},F,\sigma_F,n_\mathrm{e},\sigma_{n_\mathrm{e}},T_\mathrm{e},\sigma_{T_\mathrm{e}},C_\mathrm{TS},S_\mathrm{TS},D_\mathrm{mag},D_\mathrm{int},D_\mathrm{pol},D_\mathrm{TS},D_\mathrm{Li},D_\mathrm{equi},D_\mathrm{wall}\right)\nonumber\\
=&P\left(D_\mathrm{mag}|J_\phi,J_\mathrm{iron}\right)P\left(D_\mathrm{int}|n_\mathrm{e}\left(\psi_\mathrm{N}\right)\right)P\left(D_\mathrm{pol}|J_\phi,J_\mathrm{iron},n_\mathrm{e}\left(\psi_\mathrm{N}\right)\right)\nonumber\\
\times&P\left(D_\mathrm{TS}|n_\mathrm{e}\left(\psi_\mathrm{N}\right),T_\mathrm{e}\left(\psi_\mathrm{N}\right),C_\mathrm{TS},S_\mathrm{TS}\right)P\left(C_\mathrm{TS}\right)P\left(S_\mathrm{TS}\right)P\left(D_\mathrm{Li}|n_\mathrm{e}\left(\psi_\mathrm{N}\right),T_\mathrm{e}\left(\psi_\mathrm{N}\right)\right)\nonumber\\
\times&P\left(D_\mathrm{equi}|J_\phi,n_\mathrm{e}\left(\psi_\mathrm{N}\right),T_\mathrm{e}\left(\psi_\mathrm{N}\right),F\left(\psi_\mathrm{N}\right)\right)P\left(D_\mathrm{wall}|J_\phi,n_\mathrm{e}\left(\psi_\mathrm{N}\right),T_\mathrm{e}\left(\psi_\mathrm{N}\right)\right)P\left(J_\phi\right)P\left(J_\mathrm{iron}\right)\nonumber\\
\times&P\left(F|\sigma_{f,F},\sigma_{x,F}\right)P\left(\sigma_{f,F}\right)P\left(\sigma_{x,F}\right)P\left(n_\mathrm{e}|\sigma_{f,n_\mathrm{e}},\sigma_{x,n_\mathrm{e}}\right)P\left(\sigma_{f,n_\mathrm{e}}\right)P\left(\sigma_{x,n_\mathrm{e}}\right)\nonumber\\
\times&P\left(T_\mathrm{e}|\sigma_{f,T_\mathrm{e}},\sigma_{x,T_\mathrm{e}}\right)P\left(\sigma_{f,T_\mathrm{e}}\right)P\left(\sigma_{x,T_\mathrm{e}}\right).\label{eq:equi_joint_distribution}
\end{align}
We remark that the difference between these two models can be seen as a choice of prior knowledge. The model without the equilibrium prior takes the Gaussian process prior $J_\phi$ which forces $J_\phi$ distribution to be smooth, on the other hand, the other model takes the equilibrium prior which excludes non-equilibrium solutions.

\section{The inference}\label{sec:inference}
Given the model (joint distribution), we can calculate the posterior distribution by the Bayes' theorem. The posterior distribution for the model without the equilibrium prior is:
\begin{align}
&P\left(J_\phi,\sigma_{J_\phi},J_\mathrm{iron},n_\mathrm{e},\sigma_{n_\mathrm{e}},T_\mathrm{e},\sigma_{T_\mathrm{e}},C_\mathrm{TS},S_\mathrm{TS}|D_\mathrm{mag},D_\mathrm{int},D_\mathrm{pol},D_\mathrm{TS},D_\mathrm{Li},D_\mathrm{wall}\right)\nonumber\\
=&\frac{P\left(J_\phi,\sigma_{J_\phi},J_\mathrm{iron},n_\mathrm{e},\sigma_{n_\mathrm{e}},T_\mathrm{e},\sigma_{T_\mathrm{e}},C_\mathrm{TS},S_\mathrm{TS},D_\mathrm{mag},D_\mathrm{int},D_\mathrm{pol},D_\mathrm{TS},D_\mathrm{Li},D_\mathrm{wall}\right)}{P\left(D_\mathrm{mag},D_\mathrm{int},D_\mathrm{pol},D_\mathrm{TS},D_\mathrm{Li},D_\mathrm{wall}\right)},\label{eq:CT_posterior}
\end{align}
and with the equilibrium prior:
\begin{align}
&P\left(J_\phi,J_\mathrm{iron},F,\sigma_F,n_\mathrm{e},\sigma_{n_\mathrm{e}},T_\mathrm{e},\sigma_{T_\mathrm{e}},C_\mathrm{TS},S_\mathrm{TS}|D_\mathrm{mag},D_\mathrm{int},D_\mathrm{pol},D_\mathrm{TS},D_\mathrm{Li},D_\mathrm{equi},D_\mathrm{wall}\right)\nonumber\\
=&\frac{P\left(J_\phi,J_\mathrm{iron},F,\sigma_F,n_\mathrm{e},\sigma_{n_\mathrm{e}},T_\mathrm{e},\sigma_{T_\mathrm{e}},C_\mathrm{TS},S_\mathrm{TS},D_\mathrm{mag},D_\mathrm{int},D_\mathrm{pol},D_\mathrm{TS},D_\mathrm{Li},D_\mathrm{equi},D_\mathrm{wall}\right)}{P\left(D_\mathrm{mag},D_\mathrm{int},D_\mathrm{pol},D_\mathrm{TS},D_\mathrm{Li},D_\mathrm{equi},D_\mathrm{wall}\right)},\label{eq:equi_posterior}
\end{align}
where the denominators are a normalisation constant in this context. These posterior distributions can be explored by optimisation or sampling algorithms, for example, pattern search \cite{Hooke1961} or Markov chain Monte Carlo (MCMC) algorithms \cite{Metropolis1953, Hastings1970, Haario2001}. However, these posterior distributions are high dimensional (more than $1000$ dimensions), correlated and complex. For this reason, it is computationally challenging to explore such posterior distributions. We found a few approaches developed in the previous works \cite{Svensson2004, Ford2010, Hole2010, Hole2011, VonNessi2013, VonNessi2014}, but they did not work for this problem completely.

In this work, we developed another approach to exploring a high dimensional complex joint posterior distribution based on the Gibbs sampling scheme \cite{Geman1984}. The main idea of this approach is to separate a high dimensional joint distribution $P\left(X_1,X_2,\cdots,X_n\right)$ into a couple of low dimensional conditional distributions $P\left(X_i|X_1,\cdots,X_{i-1},X_{i+1},\cdots,X_n\right)$ and to sample them consecutively as follow:
\begin{enumerate}
	\item Begin with initial $X_1^{(k)},X_2^{(k)},\cdots,X_n^{(k)}$.
	\item Sample $X_1$ from $P\left(X_1|X_2^{(k)},X_3^{(k)},\cdots,X_n^{(k)}\right)$. Set $X_1$ to $X_1^{(k+1)}$ and sample $X_2$ from $P\left(X_2|X_1^{(k+1)},X_3^{(k)},\cdots,X_n^{(k)}\right)$. Set $X_2$ to $X_2^{(k+1)}$ and sample $X_3$ from $P\left(X_3|X_1^{(k+1)},X_2^{(k+1)},X_4^{(k)},\cdots,X_n^{(k)}\right)$. Likewise, sample all the other parameters consecutively until we get $X_1^{(k+1)},X_2^{(k+1)},\cdots,X_n^{(k+1)}$ which are the $(k+1)^\mathrm{th}$ sample.
	\item Repeat the above.
\end{enumerate}
Mathematically, these samples eventually approximate the original joint distribution $P\left(X_1,X_2,\cdots,X_n\right)$. It is usually simpler to sample each of the low dimensional conditional distributions than the high dimensional joint distribution. In addition, sometimes it is possible to break a non-linear problem into a combination of simple linear and non-linear ones in this way. In our case, we could break the full joint posterior distribution into two linear conditional distributions and one non-linear conditional distribution, therefore, reducing the difficulties of sampling substantially.

The axisymmetric plasma model without the equilibrium prior can be separated into the following parts:
\begin{align}
&P\left(J_\phi,\sigma_{J_\phi},J_\mathrm{iron}|n_\mathrm{e},\sigma_{n_\mathrm{e}},T_\mathrm{e},\sigma_{T_\mathrm{e}},C_\mathrm{TS},S_\mathrm{TS},D_\mathrm{mag},D_\mathrm{int},D_\mathrm{pol},D_\mathrm{TS},D_\mathrm{Li},D_\mathrm{wall}\right),\label{eq:CT_conditional_posterior_J}\\
&P\left(n_\mathrm{e},\sigma_{n_\mathrm{e}},T_\mathrm{e},\sigma_{T_\mathrm{e}},C_\mathrm{TS},S_\mathrm{TS}|J_\phi,\sigma_{J_\phi},J_\mathrm{iron},D_\mathrm{mag},D_\mathrm{int},D_\mathrm{pol},D_\mathrm{TS},D_\mathrm{Li},D_\mathrm{wall}\right),\label{eq:CT_conditional_posterior_p}
\end{align}
and the equilibrium model can be separated into the following parts:
\begin{align}
&P\left(J_\phi,J_\mathrm{iron}|F,\sigma_F,n_\mathrm{e},\sigma_{n_\mathrm{e}},T_\mathrm{e},\sigma_{T_\mathrm{e}},C_\mathrm{TS},S_\mathrm{TS},D_\mathrm{mag},D_\mathrm{int},D_\mathrm{pol},D_\mathrm{TS},D_\mathrm{Li},D_\mathrm{equi},D_\mathrm{wall}\right),\label{eq:conditional_posterior_J}\\
&P\left(n_\mathrm{e},\sigma_{n_\mathrm{e}},T_\mathrm{e},\sigma_{T_\mathrm{e}},C_\mathrm{TS},S_\mathrm{TS}|J_\phi,J_\mathrm{iron},F,\sigma_F,D_\mathrm{mag},D_\mathrm{int},D_\mathrm{pol},D_\mathrm{TS},D_\mathrm{Li},D_\mathrm{equi},D_\mathrm{wall}\right),\label{eq:conditional_posterior_p}\\
&P\left(F,\sigma_F|J_\phi,J_\mathrm{iron},n_\mathrm{e},\sigma_{n_\mathrm{e}},T_\mathrm{e},\sigma_{T_\mathrm{e}},C_\mathrm{TS},S_\mathrm{TS},D_\mathrm{mag},D_\mathrm{int},D_\mathrm{pol},D_\mathrm{TS},D_\mathrm{Li},D_\mathrm{equi},D_\mathrm{wall}\right).\label{eq:conditional_posterior_F}
\end{align}
Here, $\sigma_{J_\phi}$, $\sigma_{n_\mathrm{e}}$, $\sigma_{T_\mathrm{e}}$, $C_\mathrm{TS}$ and $S_\mathrm{TS}$ can be pre-optimised as follow:
\begin{enumerate}
	\item First of all, we have to get an initial guess for $\psi_\mathrm{N}$ to map the other quantities to real space. To do this, start with a Gaussian process for $J_\phi$ (after this step, for the equilibrium model, we will switch to the equilibrium prior). Given $D_\mathrm{mag}$ and $D_\mathrm{wall}$, optimise the hyperparameter $\sigma_{J_\phi}$ with the pattern search algorithm by maximising $P\left(\sigma_{J_\phi}|D_\mathrm{mag},D_\mathrm{wall}\right)$, which is proportional to the model evidence. The model evidence can be analytically calculated by the linear Gaussian inversion algorithm \cite{Svensson2008, Ford2010}. Given the optimal hyperparameter $\sigma_{J_\phi}$, infer $J_\phi$ and $J_\mathrm{iron}$ and calculate $\psi_\mathrm{N}$.
	\item\label{item:preop_repeat_from} Optimise $\sigma_{n_\mathrm{e}}$, $\sigma_{T_\mathrm{e}}$, $C_\mathrm{TS}$ and $S_\mathrm{TS}$ and then infer $n_\mathrm{e}$ and $T_\mathrm{e}$ given $\psi_\mathrm{N}$, $D_\mathrm{int}$, $D_\mathrm{TS}$, $D_\mathrm{Li}$ and $D_\mathrm{wall}$.
	\item Optimise $\sigma_F$ and then infer $F$ given $J_\phi$, $n_\mathrm{e}$, $T_\mathrm{e}$ and $D_\mathrm{equi}$ (skip this step for the model without the equilibrium prior).
	\item Update $J_\phi$, $J_\mathrm{iron}$ and $\psi_\mathrm{N}$ given all the other quantities.
	\item\label{item:preop_repeat_to} Optimise all the parameters and hyperparameters together by exploring the full joint posterior probability with the pattern search algorithm.
	\item Repeat the above from (\ref{item:preop_repeat_from}) until finding the (local) maximum.
\end{enumerate}
After this pre-optimisation, we will obtain all the parameters and hyperparameters at the (local) maximum value of the full joint posterior probability. This solution can be regarded as a maximum a~posteriori (MAP) solution and used as an initial guess for sampling. Now, we fix the hyperparameters, and then these conditional posterior distributions without the equilibrium prior can be written as:
\begin{align}
&P\left(J_\phi,J_\mathrm{iron}|\sigma_{J_\phi},n_\mathrm{e},T_\mathrm{e},C_\mathrm{TS},S_\mathrm{TS},D_\mathrm{mag},D_\mathrm{pol},D_\mathrm{wall}\right),\label{eq:CT_conditional_posterior_J_simple}\\
&P\left(n_\mathrm{e},T_\mathrm{e}|J_\phi,J_\mathrm{iron},\sigma_{n_\mathrm{e}},\sigma_{T_\mathrm{e}},C_\mathrm{TS},S_\mathrm{TS},D_\mathrm{int},D_\mathrm{pol},D_\mathrm{TS},D_\mathrm{Li},D_\mathrm{wall}\right),\label{eq:CT_conditional_posterior_p_simple}
\end{align}
and with the equilibrium prior:
\begin{align}
&P\left(J_\phi,J_\mathrm{iron}|F,n_\mathrm{e},T_\mathrm{e},C_\mathrm{TS},S_\mathrm{TS},D_\mathrm{mag},D_\mathrm{pol},D_\mathrm{equi},D_\mathrm{wall}\right),\label{eq:conditional_posterior_J_simple}\\
&P\left(n_\mathrm{e},T_\mathrm{e}|J_\phi,J_\mathrm{iron},F,\sigma_{n_\mathrm{e}},\sigma_{T_\mathrm{e}},C_\mathrm{TS},S_\mathrm{TS},D_\mathrm{int},D_\mathrm{pol},D_\mathrm{TS},D_\mathrm{Li},D_\mathrm{equi},D_\mathrm{wall}\right),\label{eq:conditional_posterior_p_simple}\\
&P\left(F|J_\phi,J_\mathrm{iron},\sigma_F,n_\mathrm{e},T_\mathrm{e},D_\mathrm{equi}\right).\label{eq:conditional_posterior_F_simple}
\end{align}
These conditional posterior distributions except the non-linear one (Equation~(\ref{eq:conditional_posterior_J_simple})) can be analytically sampled by the linear Gaussian inversion algorithm. On the other hand, the non-linear one can be sampled by the adaptive Metropolis-Hastings algorithm \cite{Metropolis1953, Hastings1970, Haario2001} with an initial proposal distribution based on its approximated analytic distribution. In the end, we repeatedly sample them in a consecutive order to collect posterior samples from the full posterior distribution.

\subsection{Inference without the equilibrium prior}\label{subsec:current_tomography}
\begin{figure}
	\centering
	\includegraphics[width=\linewidth]{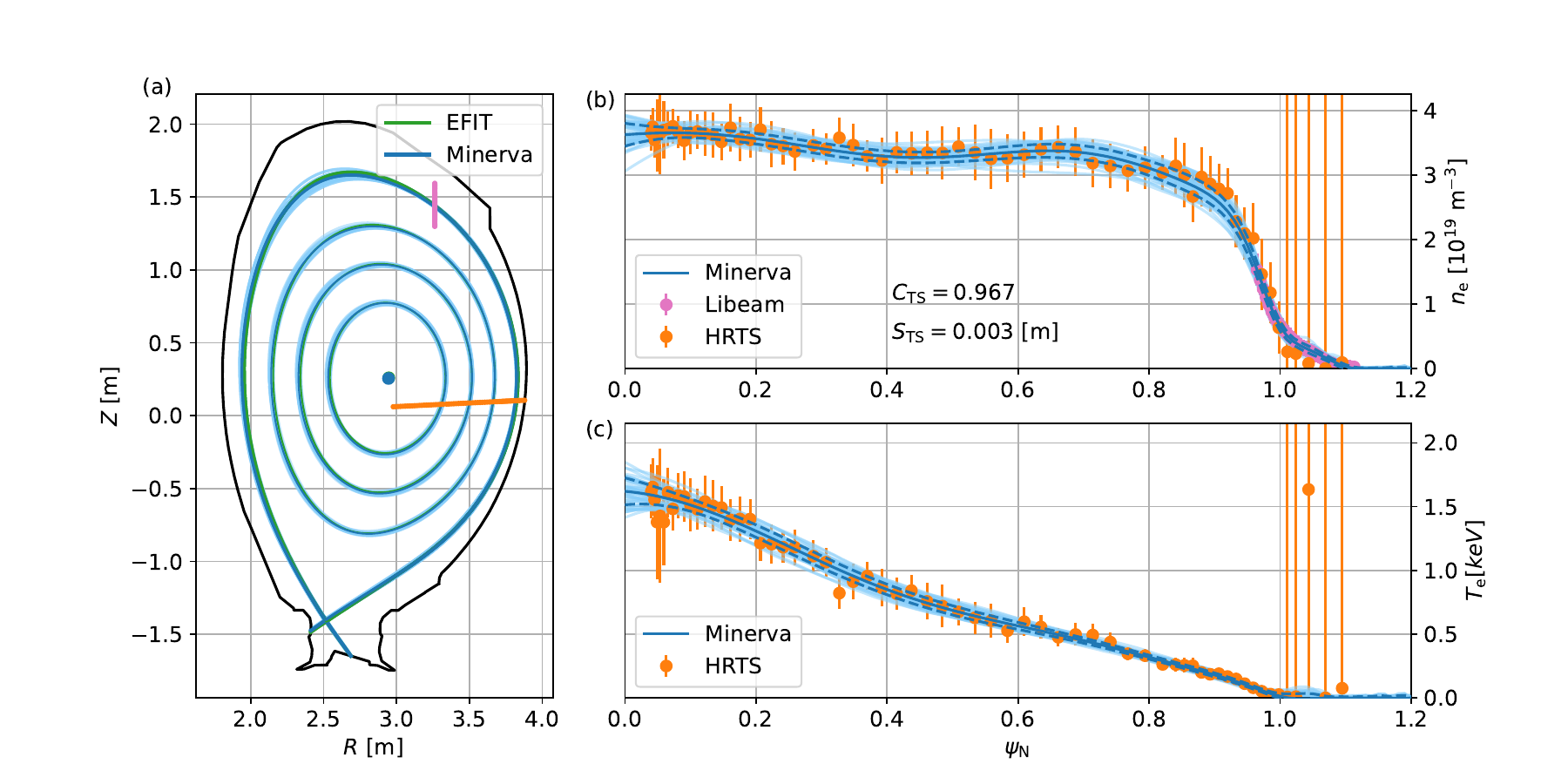}
	\caption{The results without the equilibrium prior for JET discharge \#89709 at \SI{8.0}{\second} (an L-mode plasma): (a) magnetic flux surfaces $\psi_\mathrm{N}$ on the poloidal plane, (b) $n_\mathrm{e}$ and (c) $T_\mathrm{e}$ with respect to $\psi_\mathrm{N}$. The magnetic axis, flux surfaces at $\psi_\mathrm{N}=0.25,0.50,0.75$ and the LCFS are depicted as blue dots, thin lines and thick lines. For comparison, we show the flux surfaces (in green) from the EFIT code and $n_\mathrm{e}$ and $T_\mathrm{e}$ from the conventional analysis for the HRTS (in orange) and the lithium beam (in pink) systems. The $n_\mathrm{e}$ and $T_\mathrm{e}$ positions of the HRTS and lithium beam systems are depicted as small orange and pink dots.}
	\label{fig:CT_results_L_mode}
\end{figure}

Here, we present the inferred $\psi_\mathrm{N}$, $n_\mathrm{e}$ and $T_\mathrm{e}$ without the equilibrium prior. The marginal posterior mean (in blue) and samples (in light blue) are shown in Figure~\ref{fig:CT_results_L_mode}. The magnetic axis, flux surfaces at $\psi_\mathrm{N}=0.25,0.50,0.75$ and the LCFS are depicted as blue dots, thin lines and thick lines, respectively. The first wall boundary is shown in black. The blue dashed lines are $\pm1\sigma$ posterior uncertainties for $n_\mathrm{e}$ and $T_\mathrm{e}$. For comparison, we show the flux surfaces (in green) from the EFIT code and $n_\mathrm{e}$ and $T_\mathrm{e}$ from the conventional analysis for the HRTS (in orange) and the lithium beam (in pink) systems. The $n_\mathrm{e}$ and $T_\mathrm{e}$ positions of the HRTS and lithium beam systems are depicted as small orange and pink dots. The HRTS system is automatically calibrated by inferring $C_\mathrm{TS}$ and $S_\mathrm{TS}$ given the other measurements, for example, line-integrated $n_\mathrm{e}$ from the interferometers. We note that the $n_\mathrm{e}$ values (orange dots) from the HRTS analysis are scaled with $C_\mathrm{TS}$ to avoid confusion.

%\begin{figure}
%	\centering
%	\includegraphics[width=\linewidth]{89709_53_500000_EquilibriumProfilesSimple_pub.pdf}
%	\caption{Same as Figure~\ref{fig:CT_results_L_mode} for JET discharge \#89709 at \SI{13.5}{\second} (an H-mode plasma).}
%	\label{fig:CT_results_H_mode}
%\end{figure}

\begin{figure}
	\begin{tabular}{cc}
		\includegraphics[width=.5\linewidth]{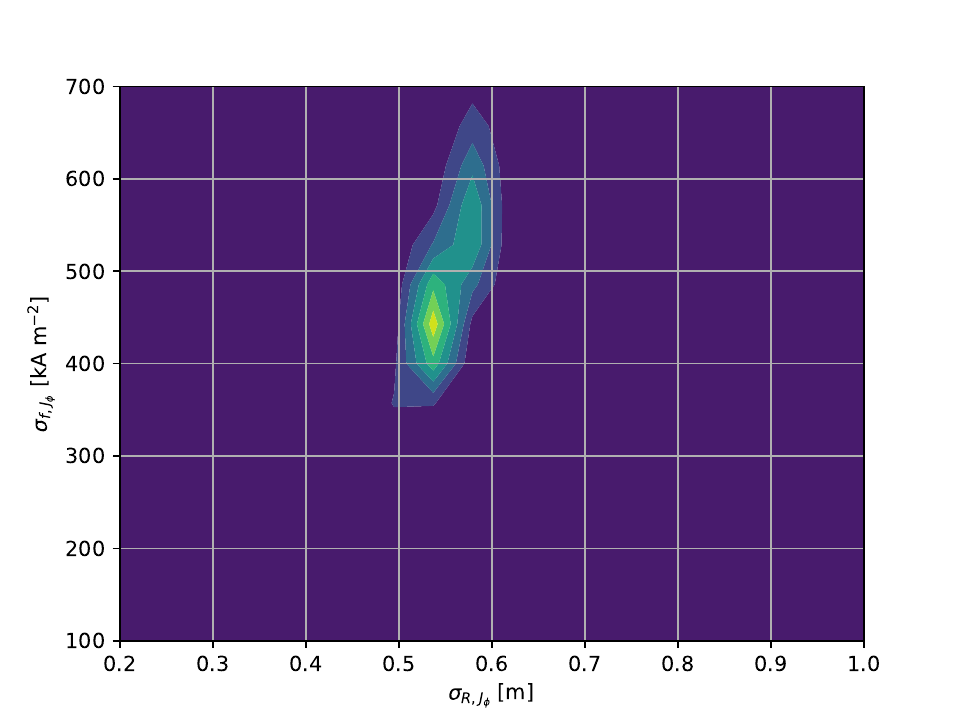} & \includegraphics[width=.5\linewidth]{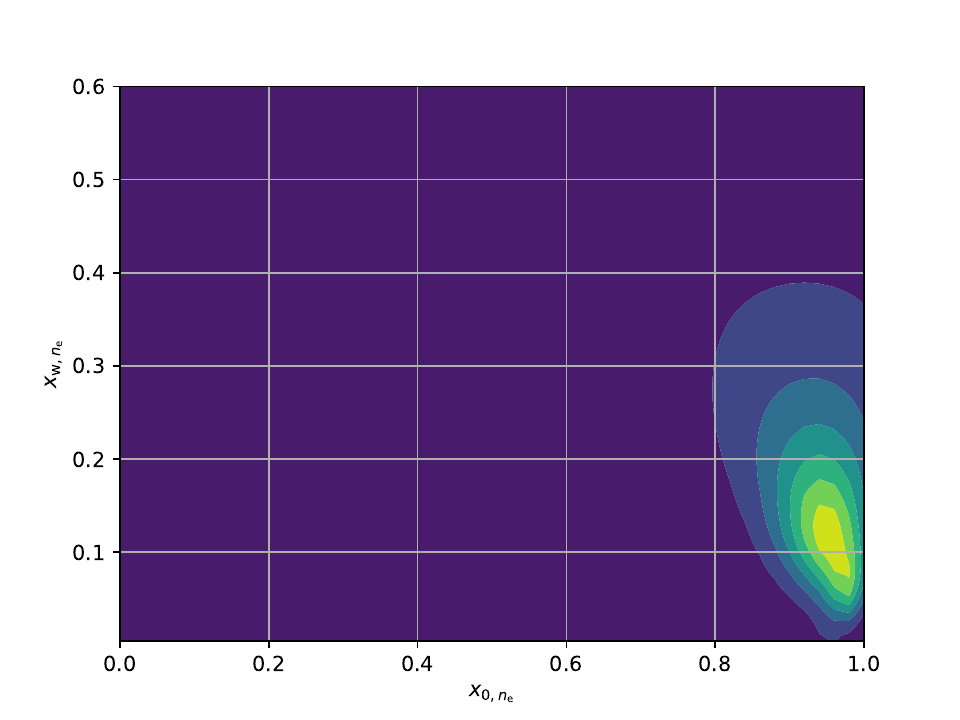} \\
		(a) $P\left(\sigma_{f,J_\phi},\sigma_{R,J_\phi}|D\right)$ & (b) $P\left(x_{w,n_\mathrm{e}},x_{0,n_\mathrm{e}}|D\right)$ \\[6pt]
	\end{tabular}
	\caption{Some examples of the posterior distributions of the hyperparameters explored during the optimisation.}
	\label{fig:evidence}
\end{figure}

The hyperparameters $\sigma_{J_\phi}$, $\sigma_{n_\mathrm{e}}$ and $\sigma_{T_\mathrm{e}}$ are optimised by maximising their posterior probabilities, which are proportional to the model evidence. Here, we show some examples of the posterior distributions of the hyperparameters in Figure~\ref{fig:evidence}. This optimisation allows us to choose the hyperparameters based on the principle of Occam's razor \cite{Gull1988, MacKay1991}. Given these \textit{optimal} hyperparameters, we can avoid under- and over-fitting. As shown in Figure~\ref{fig:CT_results_L_mode}, we can fit $n_\mathrm{e}$ and $T_\mathrm{e}$ profiles to the data in the core and edge regions without under- and over-fitting. 

\begin{figure}
	\centering
	\includegraphics[width=\linewidth]{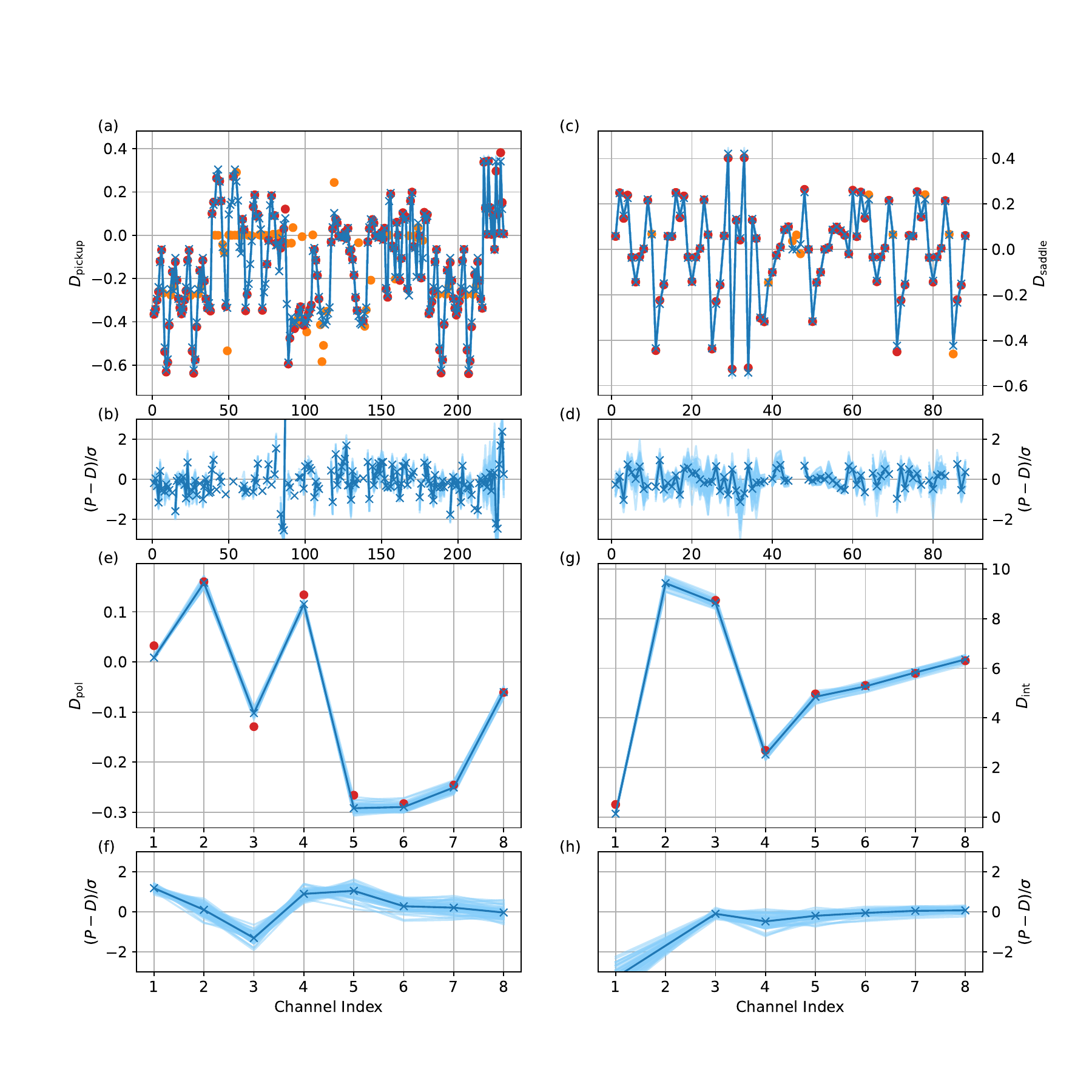}
	\caption{The predictions and observations for the (a) pickup coils, (c) saddle coils, (e) polarimeters and (g) interferometers. The predictions given the posterior mean and samples are in blue and light blue, respectively. The valid and invalid data points are in red and orange. The differences between the predictions and observations divided by their uncertainties $(P-D)/\sigma$ are calculated for the (b) pickup coils, (d) saddle coils, (f) polarimeters and (h) interferometers.}
	\label{fig:CT_predictions}
\end{figure}

The predictions and observations for the magnetic probes (pickup coils and saddle coils), polarimeters and interferometers are shown in Figure~\ref{fig:CT_predictions}. The predictions given the posterior mean and samples are in blue and light blue, respectively. Some of the magnetic probes may not be valid for some plasma discharges, due to, for example, some signal drifts over time. These invalid signals can be excluded automatically \cite{Ford2010}. Here, the valid and invalid data points are in red and orange. We also show the differences between the predictions and observations divided by their uncertainties $(P-D)/\sigma$. As shown in Figure~\ref{fig:CT_predictions}, the predictions and observations agree with each other within their predictive uncertainties. We note that the line integrated $n_\mathrm{e}$ from the second channel of the interferometer is not there for this case, nevertheless the model can still calculate the corresponding prediction.

\begin{figure}
	\centering
	\includegraphics[width=\linewidth]{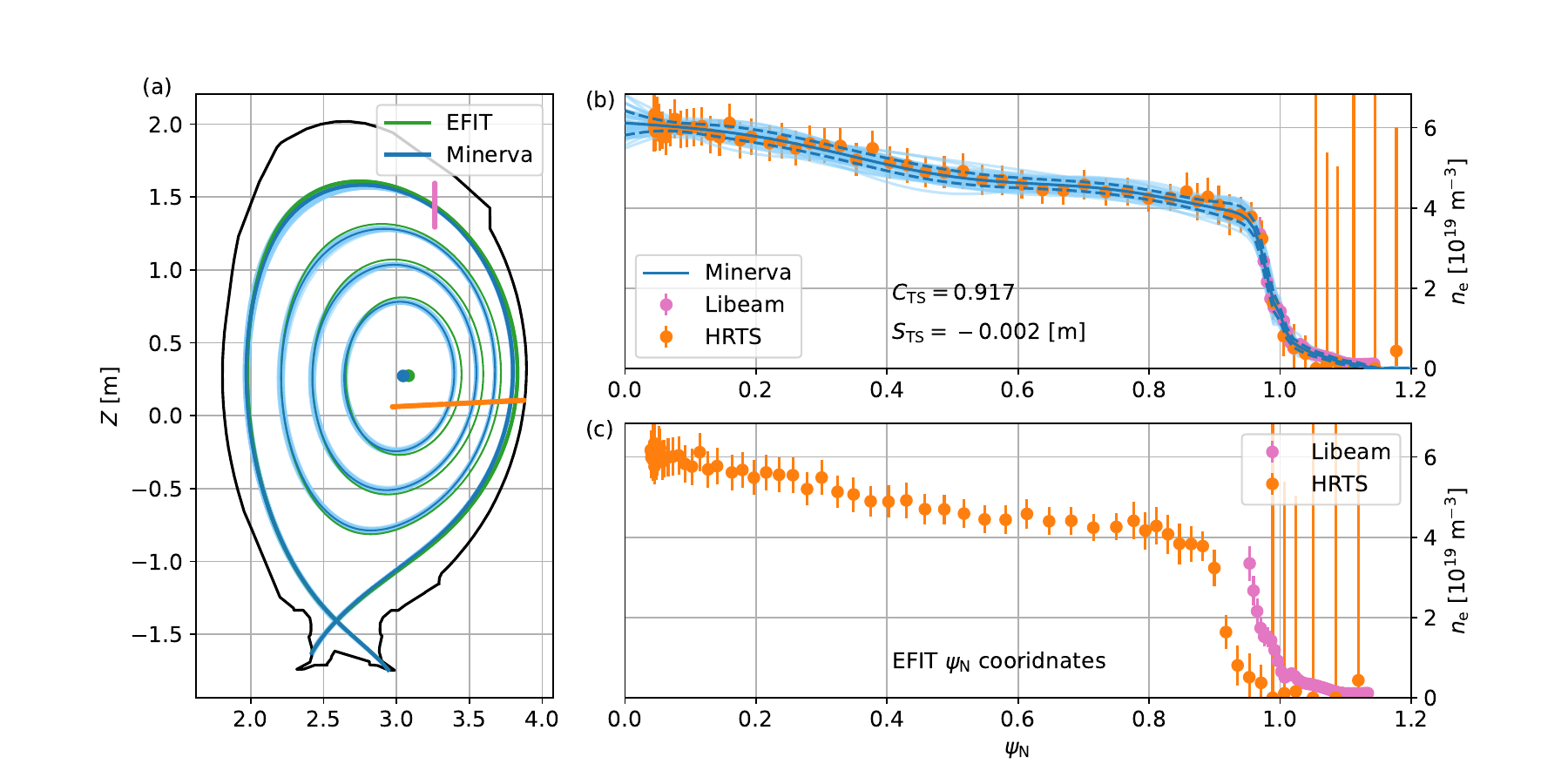}
	\caption{The inferred $n_\mathrm{e}$ profiles mapped to the magnetic flux coordinates calculated by this method and the EFIT code for JET discharge \#92398 at \SI{7.0}{\second}: (a) flux surfaces $\psi_\mathrm{N}$ from this method (in blue) and the EFIT code (in green), (b) $n_\mathrm{e}$ profiles on the flux coordinates from this method and (c) the EFIT code. The $n_\mathrm{e}$ profiles calculated by this method are consistent with both the HRTS (in orange) and lithium beam (in pink) measurements.}
	\label{fig:CT_results_comparison}
\end{figure}

\begin{figure}
	\centering
	\includegraphics[width=.7\linewidth]{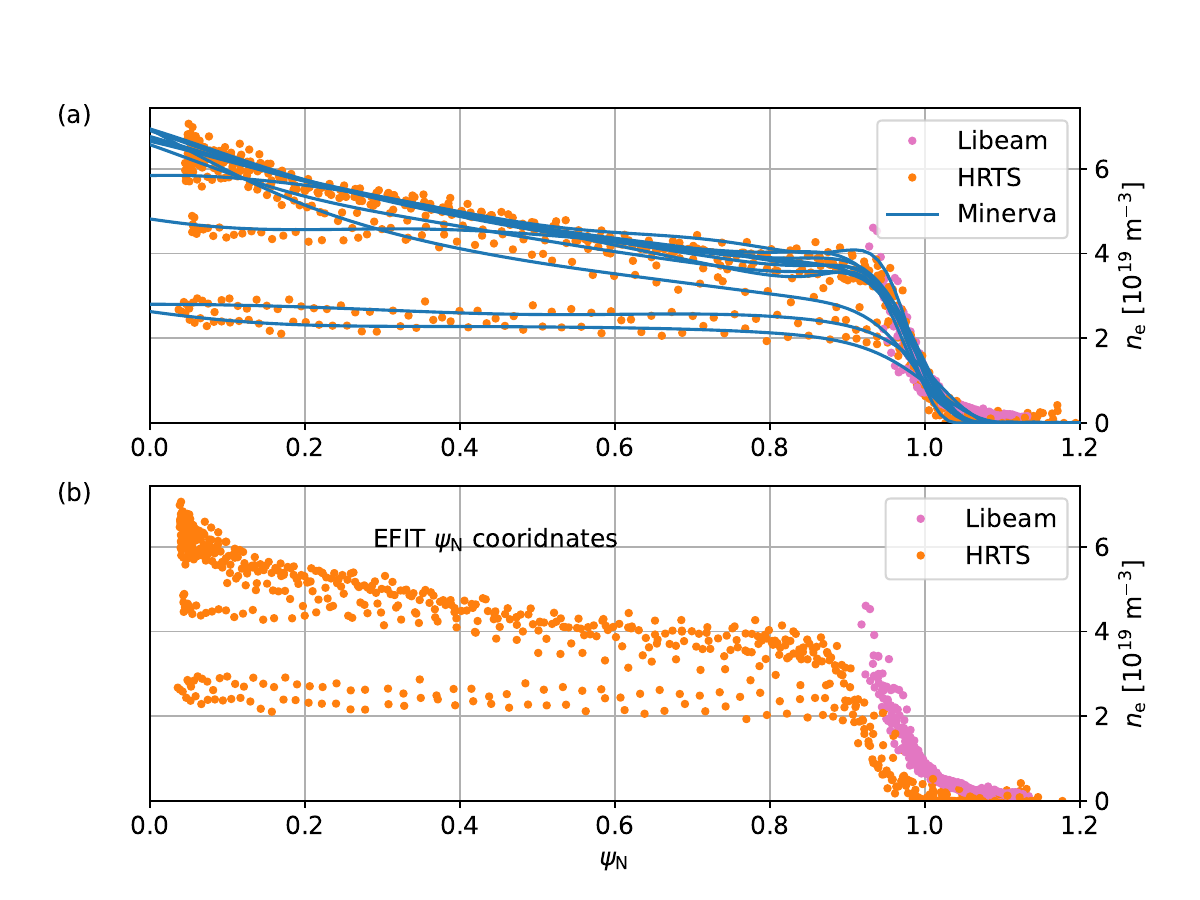}
	\caption{A time series of the inferred $n_\mathrm{e}$ profiles mapped to the $\psi_\mathrm{N}$ coordinates calculated by (a) this method and (b) the EFIT code for JET discharge \#92398 from \SI{6.0}{\second} to \SI{10.0}{\second}.}
	\label{fig:CT_results_comparison_timeseries}
\end{figure}

We remark that the inferred $\psi_\mathrm{N}$, $n_\mathrm{e}$ and $T_\mathrm{e}$ are consistent with all the measurements. Typically, the conventional analyses for individual diagnostic data map physical quantities to the flux coordinates, which are calculated by an equilibrium code such as the EFIT code. These analyses might be inconsistent with others due to not only some possible systematic inconsistencies between the diagnostics but also the flux coordinates, which might not map physical quantities in a consistent way. For example, there are two independent $n_\mathrm{e}$ measurements from the HRTS and lithium beam systems at JET which sometimes are inconsistent with each other on the EFIT $\psi_\mathrm{N}$ coordinates, as shown in Figure~\ref{fig:CT_results_comparison}(c). In such cases, it would be very difficult to figure out which data we should use for further studies. On the other hand, the method developed in this work can provide a consistent picture for all the physical quantities and measurements, as shown in Figure~\ref{fig:CT_results_comparison}(b). Moreover, since we assume some physical quantities like $n_\mathrm{e}$ and $T_\mathrm{e}$ to be constant on the same flux surface, this can indirectly provide information on $\psi_\mathrm{N}$ and $J_\phi$. Thus, the flux surfaces from this method (in blue) and the EFIT code (in orange) are different as shown in Figure~\ref{fig:CT_results_comparison}(a). On the flux surfaces from this method, the $n_\mathrm{e}$ profiles from the HRTS and lithium beam systems are consistent with each other. We also show more examples over time in Figure~\ref{fig:CT_results_comparison_timeseries}.

We emphasise that this method provides all possible solutions, which can explain all the measurements within their uncertainties. We can propagate these uncertainties to other derived physical quantities, for example, transport coefficients, to calculate their uncertainties through physics codes.

\subsection{The equilibrium inference}\label{subsec:equilibrium}
\begin{figure}
	\centering
	\includegraphics[width=\linewidth]{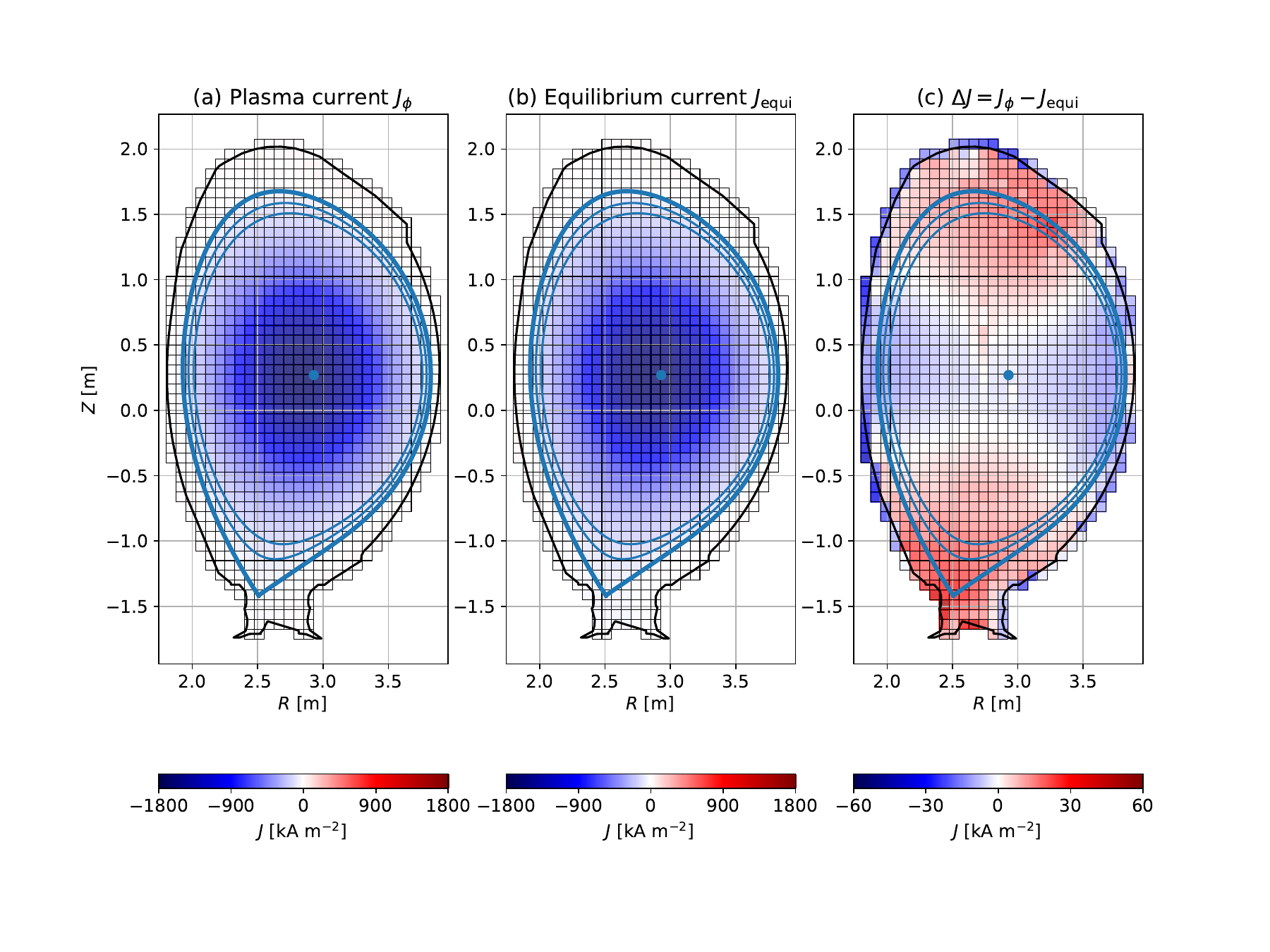}
	\caption{The equilibrium current distributions (MAP solution) for JET discharge \#89709 at \SI{8.0}{\second} (an L-mode plasma): (a) $J_\phi$ (b) $J_\mathrm{equi}=Rp^\prime+\frac{\mu_0}{R}FF^\prime$ and (c) $\Delta J=J_\phi-J_\mathrm{equi}$. The flux surfaces at $\psi_\mathrm{N} = 0.9$, $0.95$, $1.0$ are depicted as blue lines.}
	\label{fig:equi_current_L_mode}
\end{figure}

\begin{figure}
	\centering
	\includegraphics[width=\linewidth]{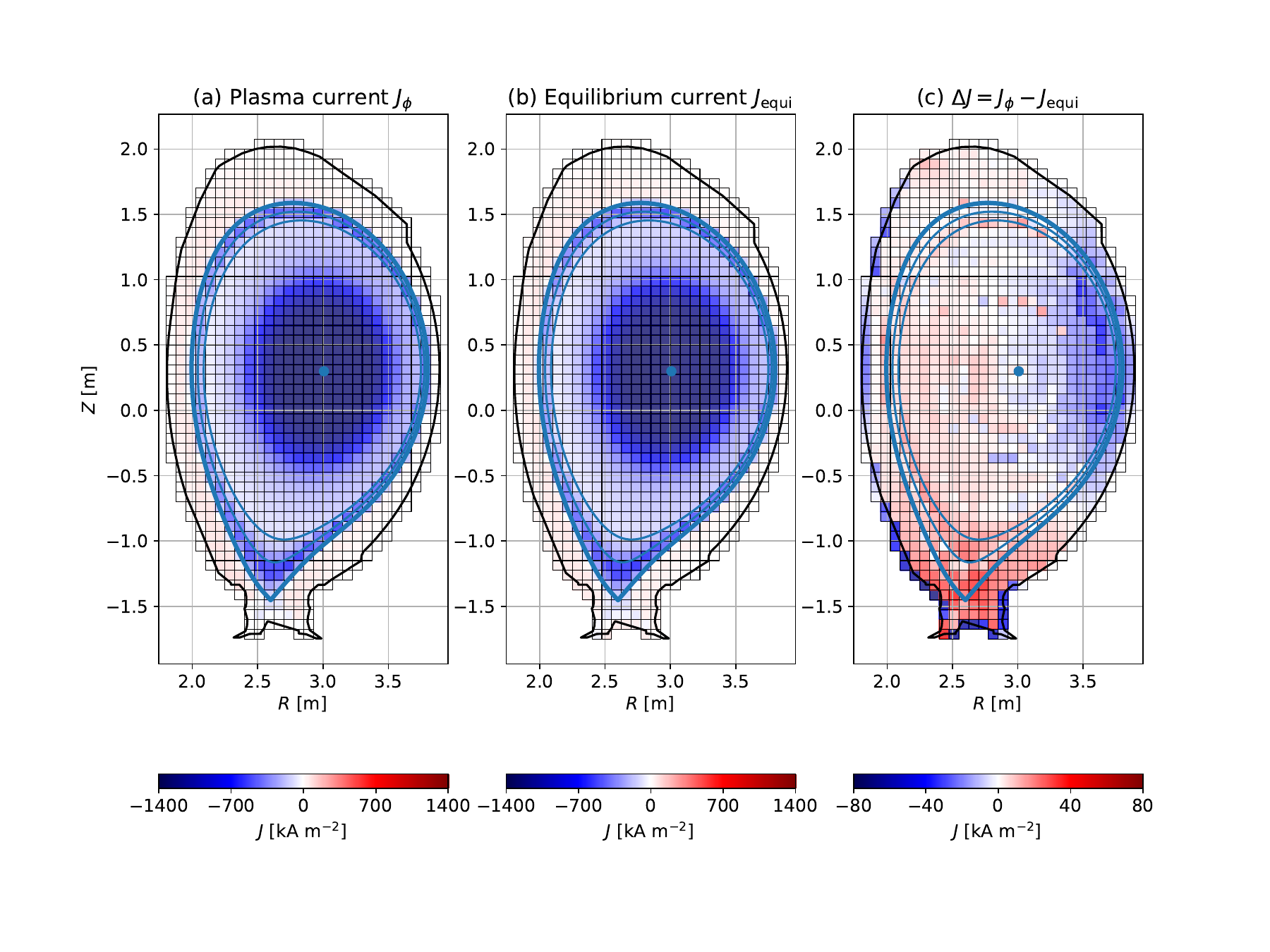}
	\caption{Same as Figure~\ref{fig:equi_current_L_mode} for JET discharge \#92398 at \SI{7.0}{\second} (an H-mode plasma).}
	\label{fig:equi_current_H_mode}
\end{figure}

Although the results without the equilibrium prior could provide a consistent picture for all the physical quantities and measurements, they might not fulfil the MHD force balance. To exclude non-equilibrium solutions, we implemented the MHD force balance constraint at every plasma current beam by introducing the virtual observations. By exploring the full joint posterior distribution with the equilibrium prior, we obtain the equilibrium current distributions for an L-mode plasma and an H-mode plasma, as shown in Figures~\ref{fig:equi_current_L_mode} and \ref{fig:equi_current_H_mode}. Here, we infer $J_\phi$, $p$ and $F$ and calculate $J_\mathrm{equi}$ given $p$ and $F$: $J_\mathrm{equi}=Rp^\prime+\frac{\mu_0}{R}FF^\prime$. The differences $\Delta J=J_\phi-J_\mathrm{equi}$ indicate that these solutions satisfy the Grad-Shafranov equation fairly well within a few per cent of the core current (less than \SI{\approx40}{\kilo\ampere}).

\begin{figure}
	\centering
	\includegraphics[width=\linewidth]{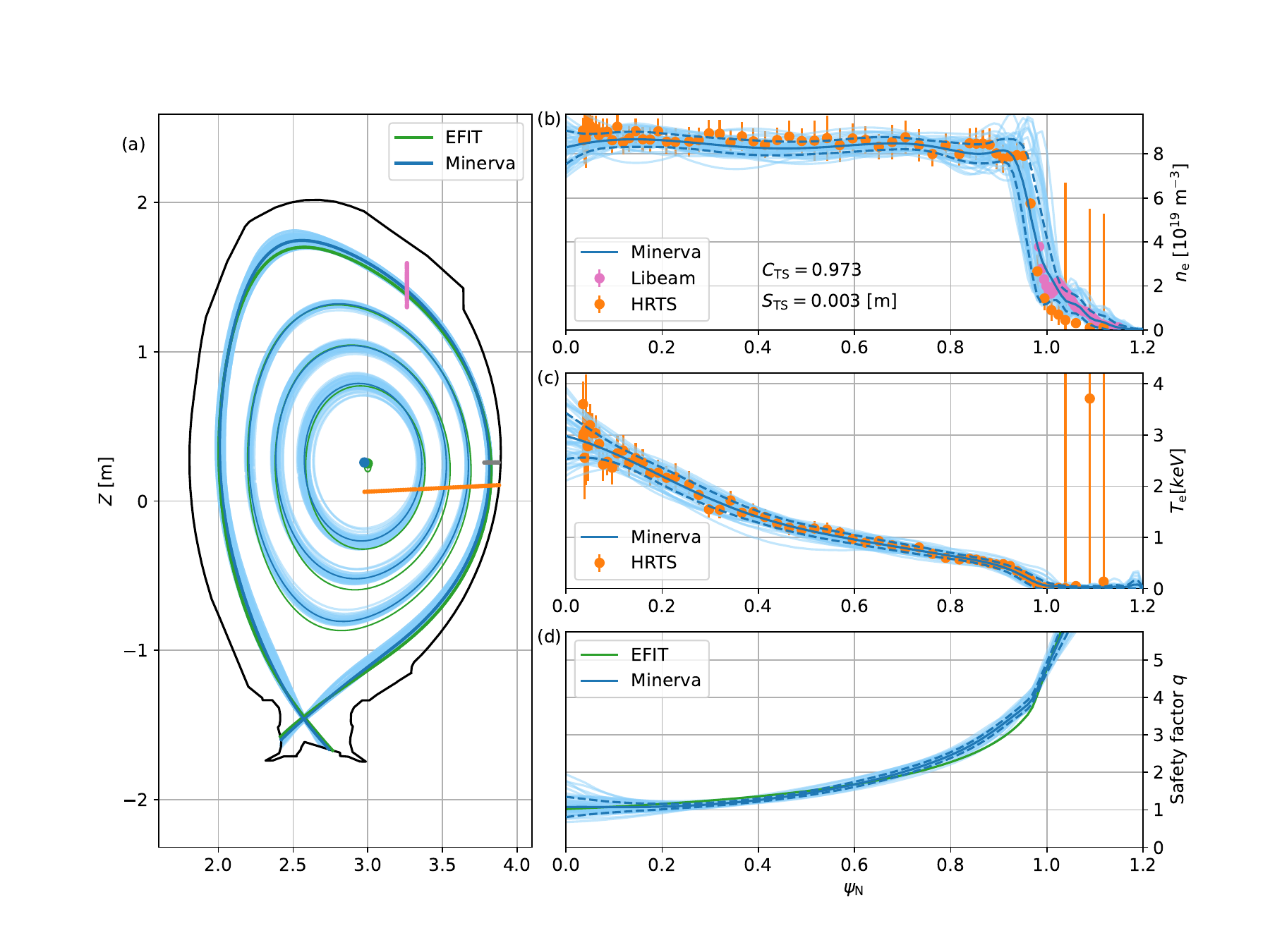}
	\caption{The results with the equilibrium prior for JET discharge \#89709 at \SI{13.5}{\second} (an H-mode plasma): (a) magnetic flux surfaces $\psi_\mathrm{N}$ on the poloidal plane, (b) $n_\mathrm{e}$, (c) $T_\mathrm{e}$ and (d) $q$ with respect to $\psi_\mathrm{N}$. The magnetic axis, flux surfaces at $\psi_\mathrm{N}=0.25,0.50,0.75$ and the LCFS are depicted as blue dots, thin lines and thick lines. For comparison, we show the flux surfaces (in green) from the EFIT code and $n_\mathrm{e}$ and $T_\mathrm{e}$ from the conventional analysis for the HRTS (in orange) and the lithium beam (in pink) systems. The $n_\mathrm{e}$ and $T_\mathrm{e}$ positions of the HRTS and lithium beam systems are depicted as small orange and pink dots.}
	\label{fig:equi_results_H_mode}
\end{figure}

We remark that the inferred $\psi_\mathrm{N}$, $n_\mathrm{e}$ and $T_\mathrm{e}$ with the equilibrium prior are also consistent with all the measurements. The marginal posterior mean (in blue) and samples (in light blue) of $\psi_\mathrm{N}$, $n_\mathrm{e}$ and $T_\mathrm{e}$ are shown in Figure~\ref{fig:equi_results_H_mode}. The $n_\mathrm{e}$ and $T_\mathrm{e}$ profiles mapped to the $\psi_\mathrm{N}$ coordinates agree with $n_\mathrm{e}$ and $T_\mathrm{e}$ from the HRTS (in orange) and lithium beam (in pink) systems. As shown in previously, $J_\mathrm{equi}$ calculated given these $n_\mathrm{e}$, $T_\mathrm{e}$ and $F$ profiles also agree with $J_\phi$. We note that this makes notable $\psi_\mathrm{N}$ between the non-equilibrium (Figure~\ref{fig:CT_results_comparison}) and equilibrium solutions (Figure~\ref{fig:equi_results_H_mode}).

Given the inferred $F$ profiles, we can calculate the safety factor $q$:
\begin{equation}
q=\frac{rB_\phi}{RB_\theta},
\label{eq:safety_factor}
\end{equation}
where $r$ is the minor radius, $B_\phi$ the toroidal magnetic field and $B_\theta$ the poloidal magnetic field. The $q$ profiles from this method and the EFIT code are similar except at the edge. We remark that the value in the core region might be determined by the Gaussian process prior, since we do not have much information at the core.

Typically, finding the MAP solution takes up to a couple of hours on a single core, but exploring the full joint posterior distribution takes much longer (approximately up to a few hundreds of hours for the equilibrium solution) due to its complexity. This can be accelerated by the machine learning approach, which is employed to speed up x-ray imaging diagnostics \cite{Pavone2018, Pavone2019}.

\subsection{Comparison between the equilibrium and non-equilibrium solutions}\label{subsec:equilibrium_comparison}
\begin{figure}
	\centering
	\includegraphics[width=\linewidth]{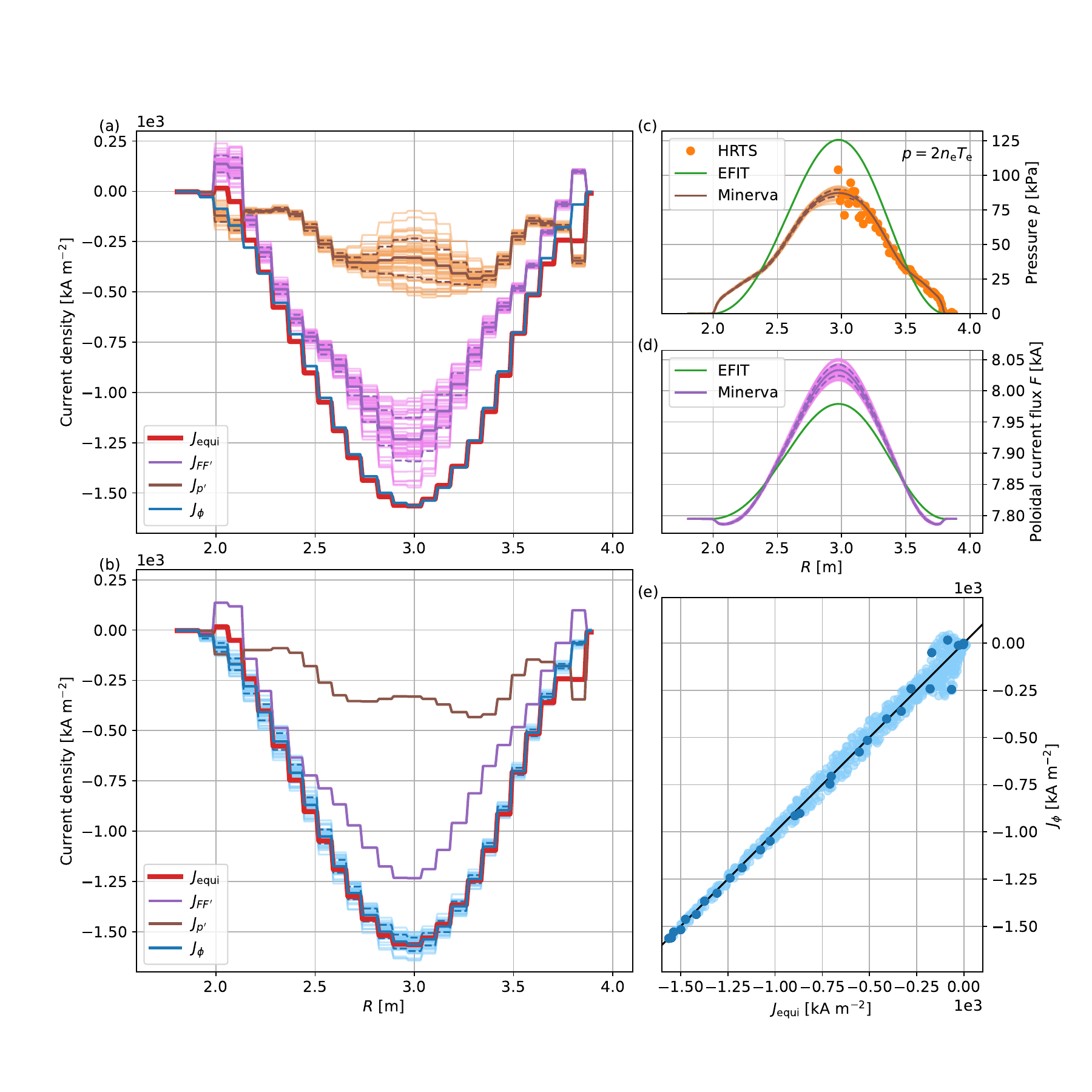}
	\caption{The equilibrium predictions given the non-equilibrium solutions for JET discharge \#89709 at \SI{13.5}{\second} (an H-mode plasma): (a) and (b) $J_\phi$ (in blue), $J_\mathrm{equi}$ (in red), $J_{p^\prime}$ (in brown) and $J_{FF^\prime}$ (c) $p$ (d) $F$ (e) $J_\phi$ and $J_\mathrm{equi}$. The posterior samples are depicted as light coloured lines. For comparison, the $p$ and $F$ profiles (in green) from the EFIT code and $p$ profiles (in orange) from the HRTS system are also shown. The black line in (e) is $y=x$.}
	\label{fig:CT_force_balance_H_mode}
\end{figure}

\begin{figure}
	\centering
	\includegraphics[width=\linewidth]{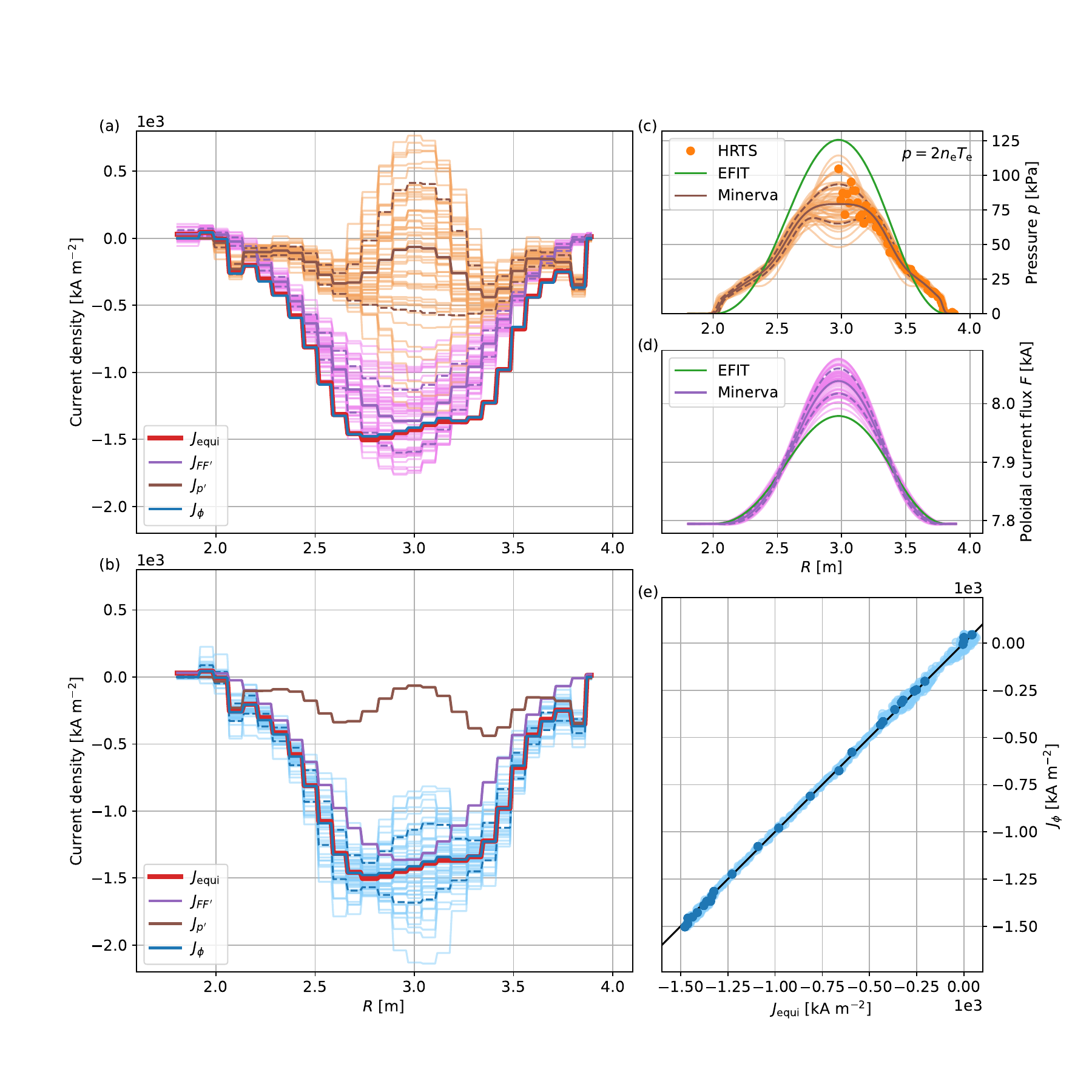}
	\caption{Same as Figure~\ref{fig:CT_force_balance_H_mode} for the inference results of the equilibrium model.}
	\label{fig:equi_force_balance_H_mode}
\end{figure}

We discussed the non-equilibrium and equilibrium solutions. Both solutions provide a consistent picture of all the physical quantities and measurements. We expect that the equilibrium solutions fulfil the MHD force balance, but not the non-equilibrium ones. Nevertheless, we can still calculate the MHD force balance predictions given these non-equilibrium solutions $\{J_\phi,J_\mathrm{iron},n_\mathrm{e},T_\mathrm{e}\}$:
\begin{align}
&P\left(F|\sigma_F,D_\mathrm{equi},\{J_\phi,J_\mathrm{iron},n_\mathrm{e},T_\mathrm{e}\}\right)\nonumber\\
=&\frac{P\left(D_\mathrm{equi}|F,\sigma_F,\{J_\phi,J_\mathrm{iron},n_\mathrm{e},T_\mathrm{e}\}\right)P\left(F|\sigma_F\right)}{P\left(D_\mathrm{equi}\right)}.\label{eq:CT_equilibrium_joint}
\end{align}
We can take the same virtual observations $D_\mathrm{equi}$ implemented in the equilibrium model and use the same Gaussian process to model $F$ profiles. Here, we show the equilibrium predictions given the non-equilibrium and equilibrium solutions in Figures~\ref{fig:CT_force_balance_H_mode} and \ref{fig:equi_force_balance_H_mode}. We select the plasma current beams across the mid-plane and present $J_\phi$ (in blue) and $J_\mathrm{equi}$ (in red) in (a) and (b). $J_\mathrm{equi}$ can be separated into $J_{p^\prime}$ (in brown) and $J_{FF^\prime}$ (in pink). The $p$ and $F$ profiles are depicted as brown and pink lines in (c) and (d). For comparison, we show the $p$ and $F$ profiles (in green) from the EFIT code and $p$ profiles (in orange) from the HRTS system. The scattered dots in (e) show differences between $J_\phi$ and $J_\mathrm{equi}$. The inferred $J_\phi$ distributions without and with the equilibrium prior are substantially different in the core and edge regions. We remark that this difference between the non-equilibrium and equilibrium solutions is due to a choice of prior knowledge. The model without the equilibrium prior predicts smooth $J_\phi$ distributions. On the other hand, the equilibrium model finds $J_\phi$ distributions satisfying the MHD force balance. Since we have a steep pressure gradient at the edge for this case, to fulfil the equilibrium condition, this should be balanced out with the Lorentz force. Given the non-equilibrium solutions, we predict $J_{FF^\prime}$ and $F$ profiles with a reversed peak at the edge, as shown in Figure~\ref{fig:CT_force_balance_H_mode}. On the contrary, the equilibrium model proposes an edge toroidal current for $J_\phi$, as shown in Figure~\ref{fig:CT_force_balance_H_mode}.

\section{Conclusions}\label{sec:conclusions}
We present a Bayesian method for inferring axisymmetric plasma equilibria consistent with magnetic field and plasma pressure measurements. The method provides all possible posterior solutions for plasma current and pressure distributions given various data from multiple plasma diagnostics including the magnetic probes, polarimeters, interferometers, high-resolution Thomson scattering and lithium beam emission spectroscopy systems. The physical quantities are modelled as Gaussian processes, and the smoothness of the processes is optimally chosen based on the principle of Occam's razor. To find equilibrium solutions, we introduce virtual observations to implement the MHD force balance constraint as a part of the prior knowledge. This equilibrium prior excludes non-equilibrium solutions in the parameter space. For comparison, we also calculate the solutions without the equilibrium prior. The high dimensional complex joint posterior distribution is explored by the new approach based on the Gibbs sampling scheme. 

The posterior solutions provide a consistent picture of all the physical quantities and measurements. The current distribution, flux surfaces, electron pressure and poloidal current flux profiles mapped on the flux coordinates are consistent with all the measurements. These solutions are self-consistent and agree with various data, thus this method could be regarded as more reliable than the analyses for individual data. Moreover, this method calculates posterior uncertainties of all these physical quantities which can be used to calculate all possible solutions for derived physical quantities, for example, transport coefficients, in further studies. 

We compared the non-equilibrium and equilibrium solutions for an H-mode plasma. Given the non-equilibrium solutions, the equilibrium constraint could be fulfilled with a reversed peak in the poloidal current flux profiles at the edge. On the other hand, the equilibrium model predicts an edge current that could balance out a steep gradient in the pedestal for an H-mode plasma. The non-equilibrium and equilibrium solutions provide different current distributions due to a choice of prior knowledge.

\section{Acknowledgement}
This work is supported by National R\&D Program through the National Research Foundation of Korea (NRF) funded by the Ministry of Science and ICT (Grant No. 2021R1A2C2005654 and 2020M1A7A03016161).

This work has been carried out within the framework of the EUROfusion Consortium, funded by the European Union via the Euratom Research and Training Programme (Grant Agreement No 101052200 — EUROfusion). Views and opinions expressed are however those of the author(s) only and do not necessarily reflect those of the European Union or the European Commission. Neither the European Union nor the European Commission can be held responsible for them.

\section*{References}
\bibliography{references}

\end{document}